\documentclass[journal]{IEEEtran}

\usepackage{cite}
\usepackage{graphicx}
\usepackage[cmex10]{amsmath}
\usepackage{algorithmicx}
\usepackage{algpseudocode}
\usepackage{algorithm}
\usepackage{dsfont}
\usepackage{color}
\usepackage{verbatim}
\graphicspath{{fig/}}

\DeclareMathOperator*{\minimize}{minimize}
\DeclareMathOperator{\subjectto}{subject~to}

\ifCLASSOPTIONcompsoc
  \usepackage[caption=false,font=normalsize,labelfont=sf,textfont=sf]{subfig}
\else
  \usepackage[caption=false,font=footnotesize]{subfig}
\fi

\algnewcommand\algorithmicinput{\textbf{INPUT:}}
\algnewcommand\INPUT{\item[\algorithmicinput]}

\algnewcommand\algorithmicoutput{\textbf{OUTPUT:}}
\algnewcommand\OUTPUT{\item[\algorithmicoutput]}

\algnewcommand\algorithmicinit{\textbf{Initialization:}}
\algnewcommand\Init{\item[\algorithmicinit]}

\algdef{SE}[DOWHILE]{Do}{doWhile}{\algorithmicdo}[1]{\algorithmicwhile\ #1}%

\newcommand{\bsl}{{\boldsymbol{l}}}
\newcommand{\Nc}{\DG{\Omega}}
\newcommand{\xx}{\boldsymbol{x}}
\newcommand{\ttv}{\boldsymbol{t}}
\newcommand{\rr}{\boldsymbol{r}}
\newcommand{\ssv}{\boldsymbol{s}}
\newcommand{\mathcalA}{A}
\newcommand{\mathcalB}{B}
\newcommand{\mathcalC}{C}

\newcommand{\mathcalM}{M}
\newcommand{\BTSs}{BTSs}

\definecolor{darkblue}{rgb}{0,0,.7}

\newcommand{\DG}[1]{{#1}}
\newcommand{\BZ}[1]{{#1}}

\makeatletter
\newenvironment{varsubequations}[1]
 {%
  \addtocounter{equation}{-1}%
  \begin{subequations}
  \renewcommand{\theparentequation}{#1}%
  \def\@currentlabel{#1}%
 }
 {%
  \end{subequations}
 }
\makeatother

\newtheorem{theorem}{Theorem}

\newtheorem{corollary}{Corollary} 

\newlength{\mywidth}
\makeatletter
\newif\ifhbonecolumn
\@ifclasswith{IEEEtran}{onecolumn}{\hbonecolumntrue}{\hbonecolumnfalse}
\makeatother
\ifhbonecolumn
\usepackage[top=1.07in, bottom=1.05in, left=1.05in,right=1.05in]{geometry}
\fi

\begin{document}
\title{Traffic-Driven Spectrum Allocation in Heterogeneous Networks}
\author{Binnan Zhuang, Dongning Guo, and Michael L. Honig\thanks{The authors are with Department of Electrical Engineering and Computer Science, Northwestern University, Evanston, IL 60208, USA.}
\thanks{The work was presented in part at Globecom 2014~\cite{ZhuGuoHon14GLOBE}. This work was supported in part by a gift from Futurewei Technologies and by the National Science Foundation under Grant No.~CCF-1018578.}
\thanks{The manuscript includes materials to appear in IEEE Journal on Selected Areas in Communications as well as an appendix omitted in the journal paper.}}

\maketitle

\begin{abstract}
Next generation cellular networks will be heterogeneous with dense deployment of small cells in order to deliver high data rate per unit area.  Traffic variations are more pronounced in a small cell, which in turn lead to more dynamic interference to other cells.  It is crucial to adapt radio resource management to traffic conditions in such a heterogeneous network (HetNet).
\DG{This paper studies the optimization of spectrum allocation in HetNets on a relatively slow timescale based on average traffic and channel conditions (typically over seconds or minutes).}
\DG{Specifically, in a cluster with $n$ base transceiver stations (\BTSs), the optimal partition of the spectrum into $2^n$ segments is determined, corresponding to all possible spectrum reuse patterns in the downlink.
Each BTS's traffic is modeled using a queue with Poisson arrivals, the service rate of which is a linear function of the combined bandwidth of all assigned spectrum segments.}
\DG{With the system average packet sojourn time as the objective, a convex optimization problem is first formulated, where it is shown that the optimal allocation divides the spectrum into at most $n$ segments.
A second, refined model is then proposed to address queue interactions due to interference, where
the corresponding optimal allocation problem admits an efficient suboptimal solution.}
\DG{Both allocation schemes attain the entire throughput region of a given network.
Simulation results show the two schemes perform similarly in the heavy-traffic regime, in which case they significantly outperform both the orthogonal allocation and the full-frequency-reuse allocation.
The refined allocation shows the best performance under all traffic conditions.}
\end{abstract}

\section{Introduction}
\label{sec:Intro}

Spectrum management in current cellular networks includes two stages: First, the locations of base transceiver stations (\BTSs) and spectrum (carrier) assignments are carefully planned offline.  Once deployed, each BTS scheduler allocates time-frequency resource blocks to users on a fast timescale.
Spectrum assignments in early networks have often been based on regular lattice frequency reuse patterns.
In current 4G networks, spectrum assignments are based either on full or fractional frequency reuse (FFR).  In FFR, a main portion of the spectrum is reused everywhere except at cell edge, and the remaining spectrum is divided for orthogonal reuse at cell edge (e.g.,~\cite{lei2007novel}).

\DG{For next generation networks, aggressive frequency reuse through dense deployment of (small) micro/pico cells is a major means for overcoming the shortage of spectrum resources~\cite{andrews2014what}.
}
\DG{Such} a heterogenous network (HetNet) with overlapping cells of all sizes often operates in the interference limited regime.  \DG{Because} small cells 
lead to more pronounced traffic \DG{and interference} variations,
traditional \DG{static} frequency reuse 
\DG{is not effective~\cite{liu2014dense}}.
\DG{Neither are the semi-regular reuse patterns of dynamic FFR
(e.g.,~\cite{
stolyar2008self-organizing,
chang2009multicell, 
ali2009dynamic}).}


\DG{In this paper, we introduce a model for a HetNet with dynamic traffic, present two optimization-based spectrum allocation schemes, and demonstrate their effectiveness using simulation.
The timescale of resource adaptation here is conceived to be relatively slow, e.g., once every a few seconds or minutes.  This timescale is, on the one hand, fast enough for tracking the aggregate traffic variation, and, on the other hand, slow enough to allow joint optimization of many cells with a large number of user equipments (UEs).
This is in contrast to most existing work, which considers resource allocation on the timescale of a frame, assuming instantaneous information exchange between cells  (see, e.g.,~\cite{
stolyar2008self-organizing,
chang2009multicell, 
ali2009dynamic,
madan2010cell,
liao2014base
}).}


\DG{Because the period of spectrum allocation is much slower than the channel coherence time, the channel conditions are accurately modeled using path loss and the statistics of small-scale fading.  Moreover, any given frequency band is assumed be homogeneous, i.e., the utility of any one Hertz of spectrum assigned to a cell depends only on the corresponding reuse pattern, i.e., the subset of cells that share it.
Specifically, in an $n$-cell HetNet, there are exactly $2^n$ distinct reuse patterns.
Indeed, the slow timescale allocation is fundamentally equivalent to deciding on the bandwidths of all those reuse patterns.
This is formulated as a convex program with optimality guarantee and relatively low computational complexity.
This is in contrast to most related work in the literature, where the allocation problem is formulated as that of deciding, for each slice of the spectrum, which \BTSs\ should use it.
The latter problem in general is a discrete optimization problem that can be hard to solve without resorting to heuristic methods, and may have many local optima (see, e.g.,~\cite{
huang2009joint,
fooladivanda2013joint,
lim2014energy-efficient,
shen2014distributed, 
deb2014algorithms
}).}

\DG{One important feature of this work is the assumption of stochastic packet arrivals to each cell.}
Recent studies
~\cite{
dhillon2013load-aware,
shojaeifard2014unified 
} point out that the usual backlogged traffic assumption
exaggerates the 
inter-cell interference in the small-cell scenario.
\DG{Another feature is the choice of delay performance as the objective for optimization.  Such a quality of service (QoS) metric is more relevant for a HetNet with dynamic traffic than the frequently used sum rate and the outage probability.}
Although there is a large body of literature on physical layer resource allocation, few papers use network layer QoS as the performance metric.
Examples include the models in~\cite{bonald2004wireless, rengarajan2008architecture}, which correspond to full-frequency-reuse (or full-reuse) only, hence spectrum allocation is not considered.
\DG{In particular,} resource allocation in~\cite{rengarajan2008architecture} is performed in the time domain by iteratively updating a scheduling policy and BTS utilizations.

The resource allocation problem here is a joint physical layer and network layer optimization problem.
To connect the spectrum (and power) resources in the physical layer to the QoS in the network layer, 
\DG{we use} the service rates of the queues \DG{of all cells} 
as the link.
\DG{Specifically,}
the spectral efficiencies along with the bandwidths of all the \DG{reuse patterns} 
allocated to a BTS determine the instantaneous service rate of the corresponding cell. The average packet delay in a given cell is in turn determined by the service rates and the packet arrival rates. Thus an optimization problem is formulated with the average packet sojourn time as the objective, the bandwidths of the \DG{reuse patterns} 
as the desired variables, and the service rates as intermediate variables.

\DG{Two allocation schemes, referred to as the ``conservative'' scheme and the ``refined'' scheme, are obtained based on different service rate models described in Section~\ref{sec:SysMod}.  The conservative scheme, discussed in Section~\ref{sec:IndQue}, assumes that a BTS's transmission rate over any spectrum segment is the worst-case rate under the corresponding reuse pattern, which is the achievable rate when all \BTSs\ in the pattern are transmitting.  In this case the queueing dynamics at one BTS does not depend on other \BTSs' activities, so that its average delay is simply that of an M/M/1 queue.
The resulting optimization problem is convex.
An important finding is that the optimal allocation uses at most $n$ out of all $2^n$ reuse patterns.  That is, it suffices to divide the spectrum into $n$ segments for allocation, rendering the solution practical.}


\DG{In the second scheme, discussed}
in Section~\ref{sec:IntQue},
\DG{the instantaneous service rate for one BTS's queue depends on whether interfering \BTSs\ are transmitting.
The analysis of $n$ interactive queues is a long-standing open problem.
To make progress, we develop an approximation for the average packet sojourn time as the objective for the optimization.  Although the problem may not be convex in general, standard solvers for convex programming appear to perform remarkably well, and the number of active reuse patterns is very close to $n$.}

\DG{Both allocation schemes are shown to}
achieve the entire throughput region of a given network.
\DG{
By simulating a cell cluster, their performance is compared with two other schemes, namely, full reuse and optimal orthogonal frequency reuse, in Section~\ref{sec:Sim}.
The refined allocation achieves the minimum delay under all circumstances.  The proposed schemes show the largest improvement in the heavy-traffic regime, where their performance is very similar.}

In independent work~\cite{KuaUtsDot2014arxiv}, Kuang et al. also considered the allocation of spectrum to an arbitrary set of reuse patterns, and incorporated user association as well.  Their performance objective is sum logarithmic utility function, so traffic and queuing models are not included in the formulation. A gradient based numerical method is used to solve the optimization problem, which has similar computational complexity to the algorithms in this paper.

\section{System Model}
\label{sec:SysMod}

\begin{figure}
\centering
\includegraphics[width=\mywidth]{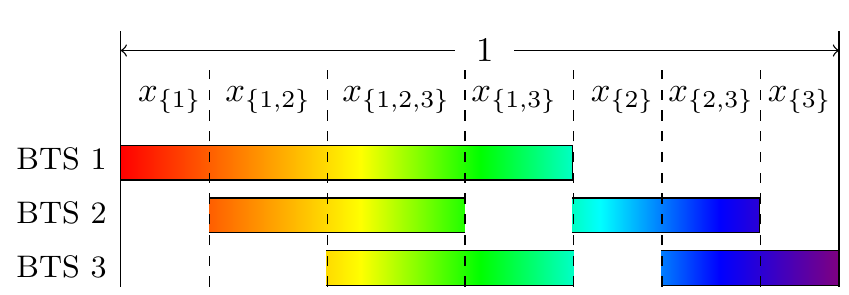}
\caption{An example of spectrum allocation to 3 \BTSs.  All $2^3-1=7$ possible nonempty reuse patterns are shown.  The bandwidth of the spectrum shared by \BTSs\ in set $A\subset\{1,2,3\}$ is denoted as $x_A$
and the total bandwidth is 1.}
\label{fig:Venn}
\end{figure}

\DG{Suppose an operator uses one licensed frequency band of $W$ Hz to carry all downlink transmissions in a HetNet of $n$ cells.}
Denote the set of \DG{$n$} 
\BTSs\ as $N=\{1,\dots,n\}$.
\DG{On a slow timescale,} 
the frequency resources are \DG{assumed to be} homogeneous.
\DG{In each period, the task of} a central controller \DG{is to} determine which part of the spectrum is allocated to each BTS.
The problem is equivalent to deciding the bandwidth \DG{of all $2^n$ reuse patterns}, 
denoted by a $2^n$-tuple:
$\xx=\left(x_\mathcalB\right)_{\mathcalB\subset N}$, where
$x_\mathcalB\in[0,1]$ is the fraction of spectrum
shared by \BTSs\ in set $\mathcalB$.
Clearly, $\sum_{\mathcalB\subset N}x_B=1$, and any efficient allocation would \DG{not use the empty reuse pattern, so that} 
$x_{\emptyset}=0$.
\DG{An example allocation to 3 \BTSs\ is shown in Fig.~\ref{fig:Venn}, where the spectrum is divided into 7 segments, corresponding to all $2^3-1=7$ nonempty reuse patterns.  The spectrum allocated to a BTS can be noncontiguous, which may be implemented using orthogonal frequency division multiplexing (OFDM).}


\DG{In this paper, we assume that either a single UE is associated with each BTS or all UEs associated with a BTS are colocated, so that they can be regarded as a single user on the slow timescale.  (This model can be generalized to allow any number of UEs at arbitrary locations, where the BTS-UE association is also to be optimized.)}
\DG{Let the aggregate traffic arriving at BTS $i$ be an independent Poisson point process with rate $\lambda_i$ packets per second.
The length of each packet independently follows the identical exponential distribution with average packet length of $L$ bits.}
\DG{Let packets intended for different UEs}
within a cell \DG{be} 
processed according to the \DG{first-in-first-out} (FIFO) criterion.\footnote{The 
results in this paper apply to all queueing disciplines that are work-conserving, non-anticipating and non-preemptive as defined in~\cite{Hav2013Springer}.}

\DG{For simplicity}, it is assumed that \DG{when} BTS $i$ \DG{transmits, it employs all reuse patterns available to it and} applies a flat power spectral density (PSD) $p_i$ over the allocated spectrum. 
At any \DG{frequency}, 
the \DG{instantaneous} spectral efficiency \DG{achievable} 
by BTS $i$ 
depends on the set of active \BTSs\ $A\subset N$ using that frequency.
Let this spectral efficiency 
be denoted by $s_{i,A}$.  Evidently, $s_{i,\mathcalA}=0$ if $i\not\in\mathcalA$. Moreover,
the spectral efficiency decreases as more \BTSs\ become active, i.e.,
$s_{i,\mathcalA}\geq s_{i,\mathcalB}$ if $i\in\mathcalA\subset\mathcalB$.
On the slow timescale, the spectral efficiencies are either known \textit{a priori} or can be computed or measured by the central controller.
\DG{For later convenience, we convert the units of $s_{i,A}$ from bits/second/Hz to packets/second by normalizing it with $L/W$ bits/packet/Hz.}
For concreteness in obtaining numerical results, we use Shannon's formula to obtain:
\begin{align}\label{eq:SpeEff}
  s_{i,\mathcalA}=
  \frac{W\, 1(i\in\mathcalA)}{L}
  \log_2\left(1+\frac{p_i} {I_{i,\mathcalA} 
    }\right)
  \quad\text{packets/second}
\end{align}
where $1(i\in\mathcalA)=1$ if $i\in\mathcalA$ and $1(i\in\mathcalA)=0$ otherwise, \DG{and} $I_{i,\mathcalA}$ is the total \DG{noise plus} interference PSD from \DG{other} \BTSs\ in $\mathcalA$ \DG{to the UEs of cell $i$, depending on the transmit PSDs and path loss.
The effect of small-scale fading can be included by considering the ergodic capacity in lieu of~\eqref{eq:SpeEff}, which does not change the main results of this paper.} 


The \DG{cells} 
form a system of $n$ (interactive) queues.
\DG{Two service rate models are conceivable:
1) In the so-called {\em conservative model}, BTS $i$ transmits at rate $s_{i,B}$ over reuse pattern $B$, which is achievable regardless of the activities of other \BTSs.
The rate contributed by reuse pattern $B$ is the product of the spectral efficiency and the bandwidth: $s_{i,B} x_B$.
Hence the service rate of cell $i$ is the sum rate of all reuse patterns, expressed as:}
\begin{align} \label{eq:r_i}
  r_i = \sum_{\mathcalB\subset N} s_{i,\mathcalB}x_\mathcalB \quad\text{packets/second}.
\end{align}
\DG{2) In the so-called {\em refined model}, every BTS adapts its rate to the instantaneous set of active \BTSs, denoted as $A$.  The rate contributed by reuse pattern $B$ to cell $i$ is 
$s_{i,B\cap A} x_B$.}
\DG{The service rate of cell $i$ depends on $A$ and is expressed as:}
\begin{align}
\label{eq:ServiceRate}
r_{i,\mathcalA}=\sum_{\mathcalB\subset N}s_{i,\mathcalB\cap\mathcalA}x_\mathcalB \quad\text{packets/second}.
\end{align}
\DG{In general, the service rate in the refined model is higher, i.e.,
$r_{i,A}\ge r_i$, because $s_{i,B\cap A}\ge s_{i,B}$.  In either case, the system is modeled as $n$ continuous-time Markov chains (CTMCs).}

\DG{In the next two sections, we develop two spectrum allocation schemes based on the preceding two service rate models, respectively.}
The objective is to minimize 
\DG{some queueing delay} by optimizing \DG{the spectrum allocation} $\xx$.

\section{A Conservative Spectrum Allocation Scheme}
\label{sec:IndQue}

\DG{In this section, the service rate model~\eqref{eq:r_i} is assumed.  This model is conservative in the sense that each BTS transmits at the worst-case rate that is achievable when all other \BTSs\ are always interfering.}
This 
is equivalent to assuming \DG{that} other 
\DG{cells' traffic is} always backlogged. 
\DG{Under this model, the cells form} 
 $n$ 
independent M/M/1 queues.
The average packet sojourn time
\DG{in cell} $i$ takes a simple form~\cite{Nel95Spinger}:
\begin{align} \label{eq:IndDelay}
  t_i=\frac{1}{r_i-\lambda_i} \quad\text{seconds}.
\end{align}
Note that $
1/(r-\lambda)$ is strictly convex in $r$ on $(\lambda,\infty)$.

\subsection{The Optimization Problem}
\label{subsec:CSAP}
The spectrum allocation problem based on the conservative approximation~\eqref{eq:IndDelay} is formulated as:
\begin{varsubequations}{P1}
\label{eq:SAP-Ind}
\begin{align}
  {\minimize\limits_{\rr,\xx}} \quad& \frac{1}{\sum_{j=1}^{n}\lambda_j}{\sum_{i=1}^n \frac{\lambda_i}{r_i-\lambda_i}}\label{eq:Obj-Ind}\\
  \subjectto\quad
  & r_i=\sum_{\mathcalB\subset N}s_{i,\mathcalB}x_{\mathcalB}, &\forall i\in N\label{eq:minServiceRate}\\
  & r_i>\lambda_i, & \forall i\in N \label{eq:Con2-Ind}\\
  & x_{\mathcalB}\geq0, &\forall \mathcalB\subset N\label{eq:Con1-Ind}\\
  & \sum_{\mathcalB\subset N}x_{\mathcalB}=1\label{eq:Con3-Ind}.
\end{align}%
\end{varsubequations}
The variables in the optimization are $\rr=[r_1,\dots,r_n]$ and $\xx$.
The objective~\eqref{eq:Obj-Ind} is the average packet delay of the entire network, where $\lambda_i/\sum_{j=1}^{n}\lambda_j$ is the fraction of total traffic
\DG{in cell} $i$. The constraints~\eqref{eq:Con2-Ind} guarantee the stability of all queues. Problem~\eqref{eq:SAP-Ind} is a convex optimization problem because
all 
constraints are linear and the objective is a linear combination of convex functions.

Since the objective 
is strictly convex and positive,
the optimization problem~\eqref{eq:SAP-Ind} has a unique global minimum when feasible%
~\cite{BerDim1999nonlinear}.
\DG{Moreover,}
due to the special structure of
\DG{the} problem, the optimal
\DG{allocation divides the spectrum into at most $n$ segments}.

\begin{theorem}\label{thm:KBTS}
  In the optimal solution of the $n$-BTS conservative spectrum allocation problem, \DG{at most $n$ out of the $2^n$ reuse patterns are active, i.e., the optimal solution $\xx$ statisfies}
  \begin{align} \label{eq:Thm}
    \left|\{\mathcalB~|~x_{\mathcalB}>0,~\mathcalB\subset N\}\right|\leq n.
  \end{align}
\end{theorem}

\begin{IEEEproof}
\label{pf:KBTS}
\DG{For every $B\subset N$, the} spectral efficiency vector 
$\ssv_{\mathcalB}=[s_{1,\mathcalB}, \dots, s_{n,\mathcalB}]$
\DG{denotes a point in $\mathds{R}^n$}. According to~\eqref{eq:minServiceRate} to \eqref{eq:Con3-Ind}, $\rr\in\mathds{R}_+^n$ is a convex combination of the $2^n$
points $(\ssv_{\mathcalB})_{\mathcalB\subset N}$ with coefficients $(x_{\mathcalB})_{\mathcalB\subset N}$, i.e., $\rr=\sum_{\mathcalB\subset N}\ssv_{\mathcalB}x_{\mathcalB}$. In other words, any $\rr$ given by~\eqref{eq:minServiceRate} is in the convex hull of $(\ssv_{\mathcalB})_{\mathcalB\subset N}$. By Carath\'eodory's Theorem~\cite{Egg69Convexity}, $\rr$ lies in a $d$-simplex with vertices in $(\ssv_{\mathcalB})_{\mathcalB\subset N}$ and $d\leq n$, i.e., $\rr$ can be written as a convex combination of $(x_{\mathcalB})_{\mathcalB\subset N}$ with at most $n+1$ nonzero \DG{coefficients in $\xx$}. 
This holds for any $\rr$ satisfying~\eqref{eq:minServiceRate} to \eqref{eq:Con3-Ind}. Furthermore, the $\rr^*$ corresponding to the optimal solution to~\eqref{eq:SAP-Ind} must be Pareto optimal in terms of the rate allocation, i.e., one cannot find another spectrum allocation $\xx$ that satisfies $r_i^*\leq \sum_{\mathcalB\subset N} s_{i,\mathcalB}x_{\mathcalB},~\forall i\in N$ with at least one of the inequalities being strict. This is because any spectrum allocation that could increase the service rate at any BTS without decreasing the rates at other \BTSs\ would also decrease the objective~\eqref{eq:Obj-Ind}. Hence $\rr^*$ cannot be an interior point of the $d$-simplex, and must lie on some $m$-face of the $d$-simplex with $m<d\leq n$. Therefore $\rr^*$ can be written as a convex combination 
\DG{with} $m+1\leq n$ nonzero \DG{coefficients in $\xx$}. 
\end{IEEEproof}

\begin{corollary}\label{cor:Sub}
  In the optimal solution of the $n$-BTS conservative spectrum allocation problem, for any subset $\mathcalM\subset N$ of 
$m$ \BTSs, the spectrum exclusively used by those \BTSs\ in $\mathcalM$ is divided into at most $m$ segments:
\begin{align}
\label{eq:Cor}
\left|\{\mathcalB~|~x_B>0,~\mathcalB\subset\mathcalM\}\right|\leq m.
\end{align}
\end{corollary}
\begin{IEEEproof}
\label{pf:sub}
If
\DG{$x_B$ for every $B\subset N$, $B\not\subset M$ is fixed (at its optimal value)},
 then~\eqref{eq:SAP-Ind} becomes an optimization problem over variables $(x_{\mathcalB})_{\mathcalB\subset\mathcalM}$. The service rates at \BTSs\ not in $\mathcalM$ are fixed, and the service rates for the $m$ \BTSs\ in $\mathcalM$ are convex combinations of $(x_{\mathcalB})_{\mathcalB\subset\mathcalM}$ plus a constant vector in $\mathds{R}_+^m $. The optimization problem reduces to the form of (P1) with only $m$ \BTSs. Hence Corollary~\ref{cor:Sub} is proved using the same arguments used to prove Theorem~\ref{thm:KBTS}.
\end{IEEEproof}


\subsection{An Efficient Algorithm}
\label{subsec:EA}

\DG{The optimization problem~\eqref{eq:SAP-Ind} has $n+2^n$ variables.}
The computational complexity is typically polynomial in $2^n$ using a standard convex optimization \DG{solver}. 
The structure of the optimal solution given by Theorem~\ref{thm:KBTS} suggests \DG{a more efficient solution, as}
we only need to \DG{determine} 
the \DG{bandwidths} 
of the $n$ nonzero segments. The difficulty \DG{of course} is to decide which $n$ segments. Algorithm~\ref{alg:SAP-Ind} 
\DG{solves} the $n$-BTS spectrum allocation problem \DG{iteratively, the details of which are explained next}.

\begin{algorithm}
\caption{
The conservative spectrum allocation \DG{scheme}
}
\label{alg:SAP-Ind}
\begin{algorithmic}[]
\INPUT {$\lambda_i$ and $s_{i,\mathcalB}$ for all $i\in  N$ and $\mathcalB\subset N$.}

\OUTPUT{$(x_{\mathcalB})_{\mathcalB\subset N}$}.

\Init{Find a feasible solution $(x'_{\mathcalB})_{B\subset N}$ by solving~\eqref{eq:SAP-Ind} with constant objective.
$\Nc \leftarrow \{\mathcalB~|~x'_{\mathcalB}>0\}$, $\Nc'\leftarrow\emptyset$.}

\While{$\Nc\not\subset \Nc'$}
    \State 1. $\Nc' \leftarrow \Nc$;
    \State 2. Find $(x_{\mathcalB})_{\mathcalB\subset N}$ by solving~\eqref{eq:SAP-Ind} starting from $(x'_{\mathcalB})_{\mathcalB\subset N}$ with 
    additional constraints: $x_B=0,~\forall\mathcalB\notin \Nc$;
    \State 3. Compute the partial derivatives of the objective function~\eqref{eq:Obj-Ind} with respect to each element in $(x_B)_{B\subset N}$, namely, $\Delta_{x_B} \leftarrow -\sum_{i\in\mathcalB}\frac{\lambda_is_{i,\mathcalB}}{(r_i-\lambda_i)^2}$;
    \State 4. $\Nc \leftarrow \{\mathcalB~\text{for}~\text{the}~n~\text{smallest}~\Delta_{x_B}\}$, $\Nc \leftarrow \Nc\cup \Nc'$, $(x'_{\mathcalB})_{\mathcalB\subset N} \leftarrow (x_{\mathcalB})_{\mathcalB\subset N}$.
\EndWhile
\end{algorithmic}
\end{algorithm}

\subsubsection{Initialization}
\label{subsubsec:Init}

To find a feasible point to start, we first
solve a modified
~\eqref{eq:SAP-Ind} by replacing the objective function~\eqref{eq:Obj-Ind} with a constant.
The 
problem can be transformed to a linear program in standard form, which can be solved using the simplex method~\cite{BerDim97LP}. Although the worst-case complexity of this method is $O(2^{2^n})$, for most practical problems the complexity is usually
\DG{very low}.
According to the properties of basic feasible solution to a linear program~\cite{BerDim97LP}, the solution (an initial point for~\eqref{eq:SAP-Ind}) will have at most $n+1$ \DG{active reuse patterns, which are collected in the {\em set of candidate reuse patterns}, denoted as $\Nc$.}


\subsubsection{Description \DG{of the Iterations}}
\label{subsubsec:Alg}
The main idea is similar to the delayed column generation algorithm for solving large-scale linear programs. The difference 
is the criterion used to
\DG{select variables to be added to the set of candidate reuse patterns, $\Nc$.} 
In each iteration, the algorithm \DG{first} finds the optimal solution within the candidate set $\Nc$.
\DG{Then,} the partial derivatives with respect to 
$x_B$ \DG{for all $B\subset N$} 
are calculated (including those not in $\Nc$). The
$n$
\DG{reuse patterns}
with the 
smallest derivatives are added to the candidate set \DG{$\Nc$, to be used in the next iteration}.  (The number of variables added to the candidate set may be \DG{fewer} 
than $n$ \DG{due to overlap}.)
The algorithm \DG{terminates} 
when the candidate set \DG{ceases to grow}. 
The proposed algorithm is guaranteed to converge to the global optimum, \DG{because otherwise it can always find at least one new reuse pattern to add to $\Nc$}.
In the worst-case, the candidate set 
\DG{may} eventually include all $2^n$ variables, \DG{but typically the algorithms terminates quickly}.


\subsubsection{Performance}

\begin{figure}
\centering
\includegraphics[width=\mywidth]{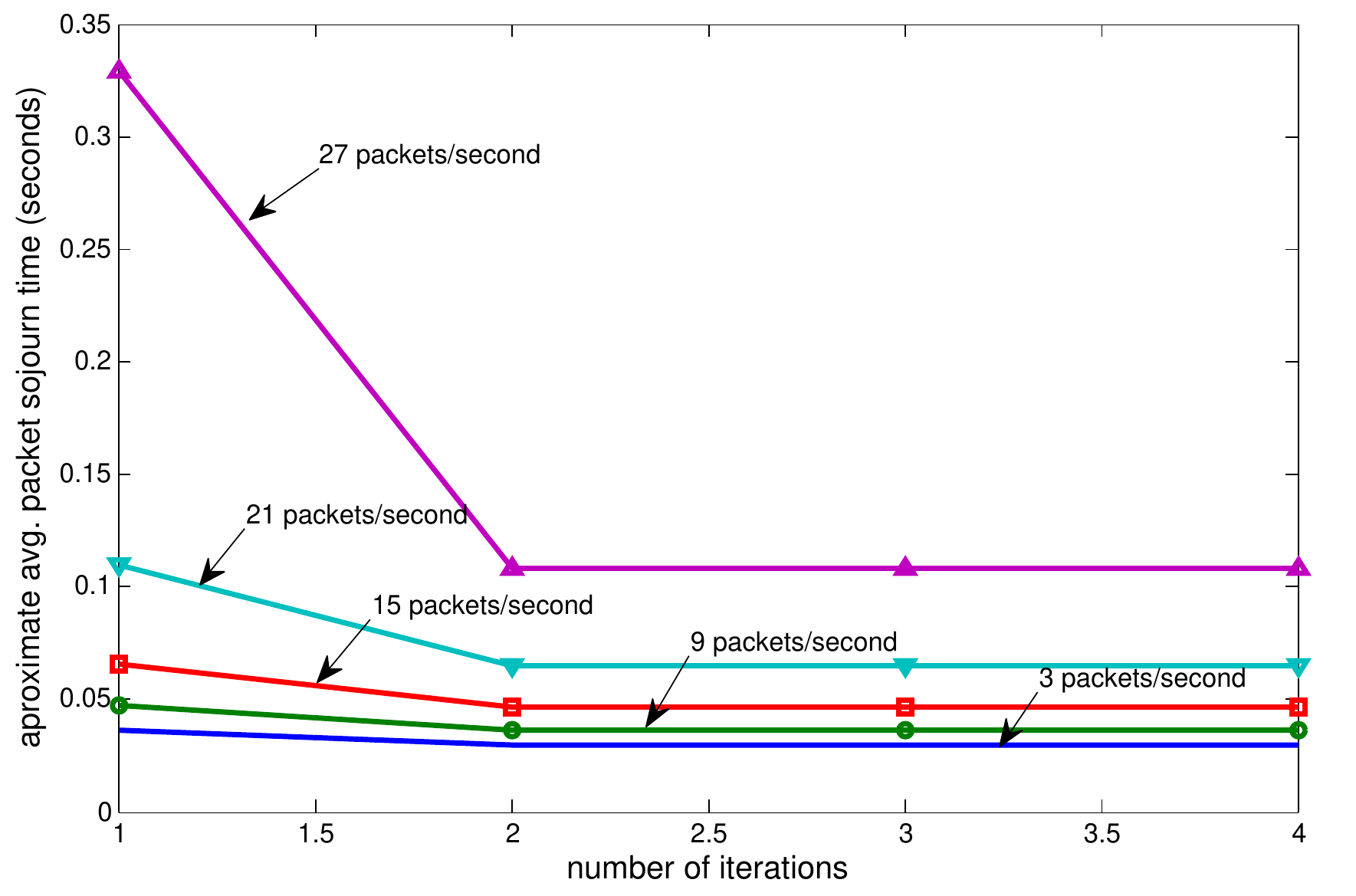}
\caption{Approximated average 
delay versus number of iterations using Algorithm~\ref{alg:SAP-Ind} with different average packet arrival rates.}
\label{fig:Alg}
\end{figure}

\DG{Usually, the fewer reuse patterns the algorithm begins with, the faster it converges.}
\DG{Thus if} the full-reuse allocation ($x_N=1$) \DG{is feasible, it is a preferred} 
initial point.  
\DG{At any rate, the initial solution} 
has no more than $n+1$ \DG{active reuse patterns}. 
Example \DG{plots} of delay versus \DG{the} number of iterations are shown in Fig.~\ref{fig:Alg} with
7 cells and different traffic loads.
In the simulation, Algorithm 1 starts with the full-reuse allocation
\DG{and} converges to the global optimum within a few iterations.

\section{A Refined Spectrum Allocation Scheme}
\label{sec:IntQue}

In this section, \DG{we adopt the refined service rate model~\eqref{eq:ServiceRate} and develop a corresponding spectrum allocation scheme.}
We assume each BTS adapts its transmit rate 
to 
\DG{the instantaneous} interference \DG{level at its UEs}.
One way to implement this in practice without explicit
knowledge of the interference levels 
is to use a hybrid automatic repeat request (HARQ) scheme.

The $n$ \DG{cells} 
form a system of $n$ (interactive) queues, where the instantaneous service rate of each queue depends on which other queues are empty.
\DG{Under such} a \emph{coupled-processors} model,
\DG{it is challenging to express the queueing delay as a function of the spectrum allocation and the arrival rates.}
In the special case of two coupled queues, finding the joint steady-state distribution can be formulated as a Riemann-Hilbert problem~\cite{FayIas1979Springer}. The same two-dimensional Markov process has been studied in~\cite{AdaWes1993AAP}, where the steady-state distribution can be represented as an infinite series of product forms. Two coupled processors with generally distributed service times have been studied in~\cite{CohBox2000Elsevier}, which shows the joint workload distribution can be determined by solving a boundary value problem. These results are difficult to use for numerical computation. Also, few results exist for more than two coupled queues.

\DG{In Section~\ref{subsec:RSAP},}
we derive an 
approximation for the \DG{average packet sojourn time based on a modified queueing system, where the queue interactions are somewhat simplified}.
\DG{A spectrum allocation scheme is then developed in Section~\ref{s:opt2} to minimize the delay approximation.}
\DG{In Section~\ref{subsec:Pre}, we visit}
the best known upper and lower bounds on the
\DG{actual} 
delay.
It is \DG{then} shown in Section~\ref{sec:Stable} that the \DG{proposed} 
approximation is between existing upper and lower bounds.

\subsection{\DG{Delay Approximation}} 
\label{subsec:RSAP}

\begin{figure}
\centering
\includegraphics[width=3in]{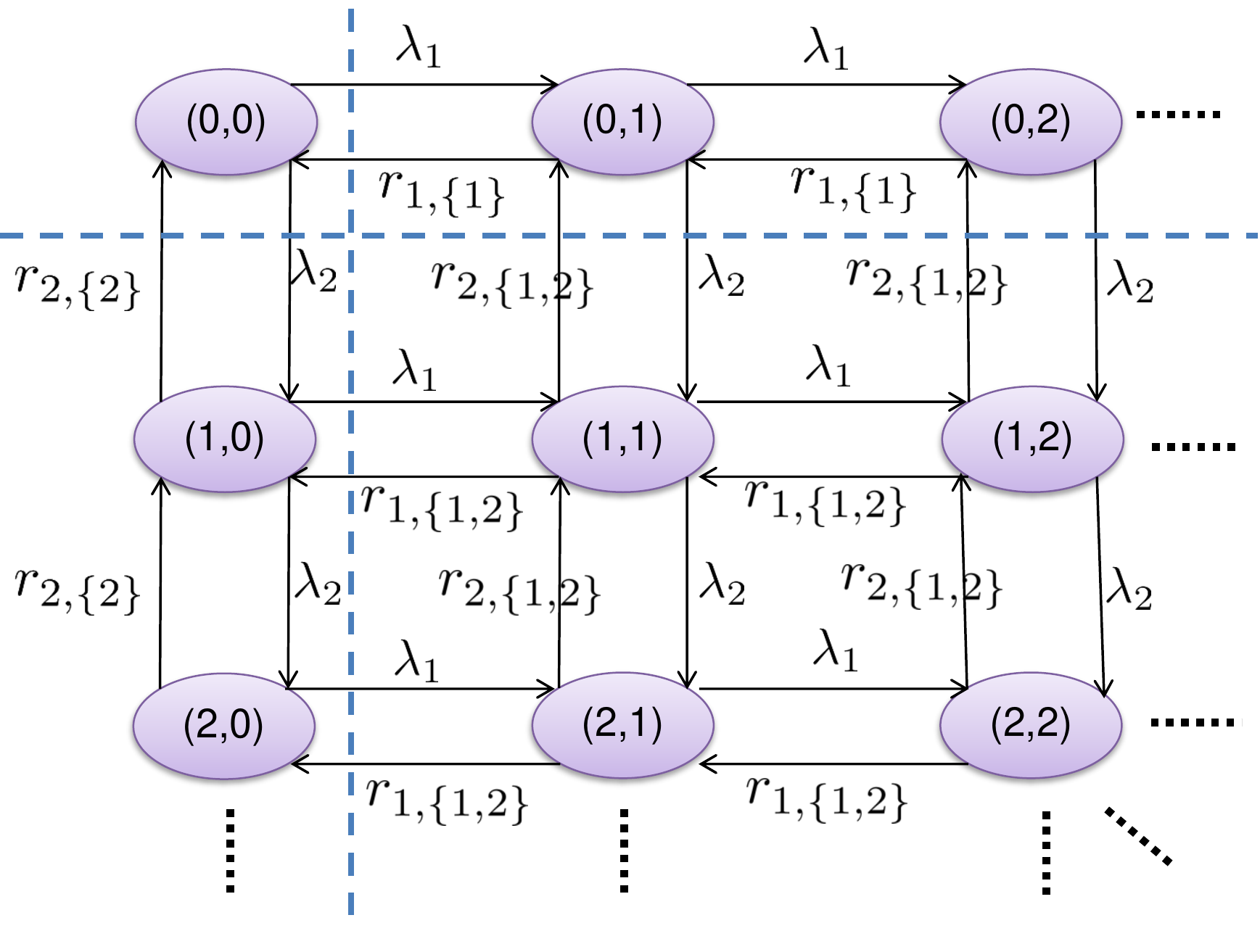}
\caption{The continuous time Markov Chain for the 2-BTS interactive queueing model.
\label{fig:MarkovChain}}
\end{figure}

The original $n$-dimensional CTMC can be described as follows.
Let $\bsl=[l_1,\dots,l_n]$ be the state of the CTMC, where $l_i\ge0$ is the number of packets in queue $i$ (including the packet being served, if any).
Let $\mathcalA_\bsl=\{i\,|\,l_i>0\}$ denote the corresponding set of active \BTSs.
The transition rate from state $\bsl$ to state $\bsl'=[l'_1,\dots,l'_n]$ is given by:
\begin{align}\label{eq:Prob}
Q(\bsl,\bsl')=
\begin{cases}
\lambda_i,&\text{if}~l_i+1=l'_i,~l_j=l'_j~\forall j\neq i\\
r_{i,\mathcalA_\bsl},&\text{if}~l_i-1=l'_i,~l_j=l'_j~\forall j\neq i\\
-\underset{\bsl'':\bsl''\neq \bsl}{\sum} Q(\bsl,\bsl''),&\text{if}~\bsl=\bsl'\\
0,&\text{otherwise}
\end{cases}
\end{align}
where the service rate $r_{i,\mathcalA}$ 
is defined in~\eqref{eq:ServiceRate}. 
\DG{As an example,} the CTMC for a 2-BTS system is illustrated in Fig.~\ref{fig:MarkovChain}.

\DG{To ease the analysis,} we shall approximate the original CTMC by a modified
CTMC with reduced memory.
Because the service rate $r_i$ depends only on the set of active queues, we group all the states corresponding to \DG{each} 
active set $\mathcalA$ and refer to them as group $A$.
According to~\eqref{eq:Prob}, transitions occur between neighboring states where the length of a single queue increases or decreases by 1.
Moreover, the CTMC makes a certain number of transitions between states within a group $A$ before it jumps into a different group $A'$.
Specifically, we make the following two (somewhat strong) assumptions about the modified CTMC:
\begin{enumerate}
\item When the modified CTMC transits from some state $\bsl$ in group $A$ to some random state $\bsl'$ in a different group $A'$, the new state 
is independent of $\bsl$.  
\DG{In other words,} such a transition is memoryless;
\item When the modified CTMC transits from group $A$ to a different group $A'$,
the probability of assuming any state in $A'$ is proportional to the stationary distribution of the state in the modified CTMC.
\end{enumerate}

In the following, we analyze the stationary distribution of the modified CTMC.
By definition, each inter-group state transition of the CTMC is a renewal, because the state is chosen anew within the new group according to the steady-state distribution.
In addition, the intra-group transitions within a group form an independent CTMC.
An important observation is that, within a group, where the set of active \BTSs\ is fixed, all the service rates are invariant with the queue lengths, so that the $n$ queues become independent.
The probability of a state $\bsl$ is thus decomposed as
\begin{align}  \label{eq:ProbJoint}
  p(\bsl) = p(A_\bsl) \prod^n_{i=1} p_i(l_i|A_\bsl)
\end{align}
where $p_i(l|A)$ is the probability that queue $i$ has length $l$ given that the state is in group $A$.
Evidently, $p_i(0|A)=0$ if $i\in A$,
$p_i(0|A)=1$ if $i\notin A$,
and
$p_i(l|A)=0$ if $i \notin A,~l>0$.
It suffices to determine the steady-state distribution of the groups $A\subset N$,
and, for each group $A$, the steady-state distribution of the states in $A$.

\subsubsection{The Intra-group CTMC}





Consider the independent CTMC within any given group $A$.
The CTMC can only exit group $A$ when some queue length is equal to 1.
For queue $i\in A$, the probabilities of states 1, 2, $\dots$ must satisfy the detailed balance equation:
\begin{align}\label{eq:DetailBE}
  p_{i}(l|\mathcalA)\lambda_i = p_{i}(l+1|\mathcalA)r_{i,\mathcalA}, \quad l=1,2,\dots.
\end{align}
As a result, for every $i\in A$,
\begin{align}  \label{eq:3}
  p_i(l|A)
  =
  \left( 1 - \frac{\lambda_i}{r_{i,\mathcalA}} \right)
  \left(\frac{\lambda_i}{r_{i,\mathcalA}}\right)^{l-1}, \quad l=1,2,\dots.
\end{align}
It is easy to check that
\begin{align}  \label{eq:4}
  \sum^\infty_{l=1} p_i(l|A) = 1.
\end{align}


\subsubsection{The Inter-group CTMC}

\begin{figure}
\centering
\includegraphics[width=3in]{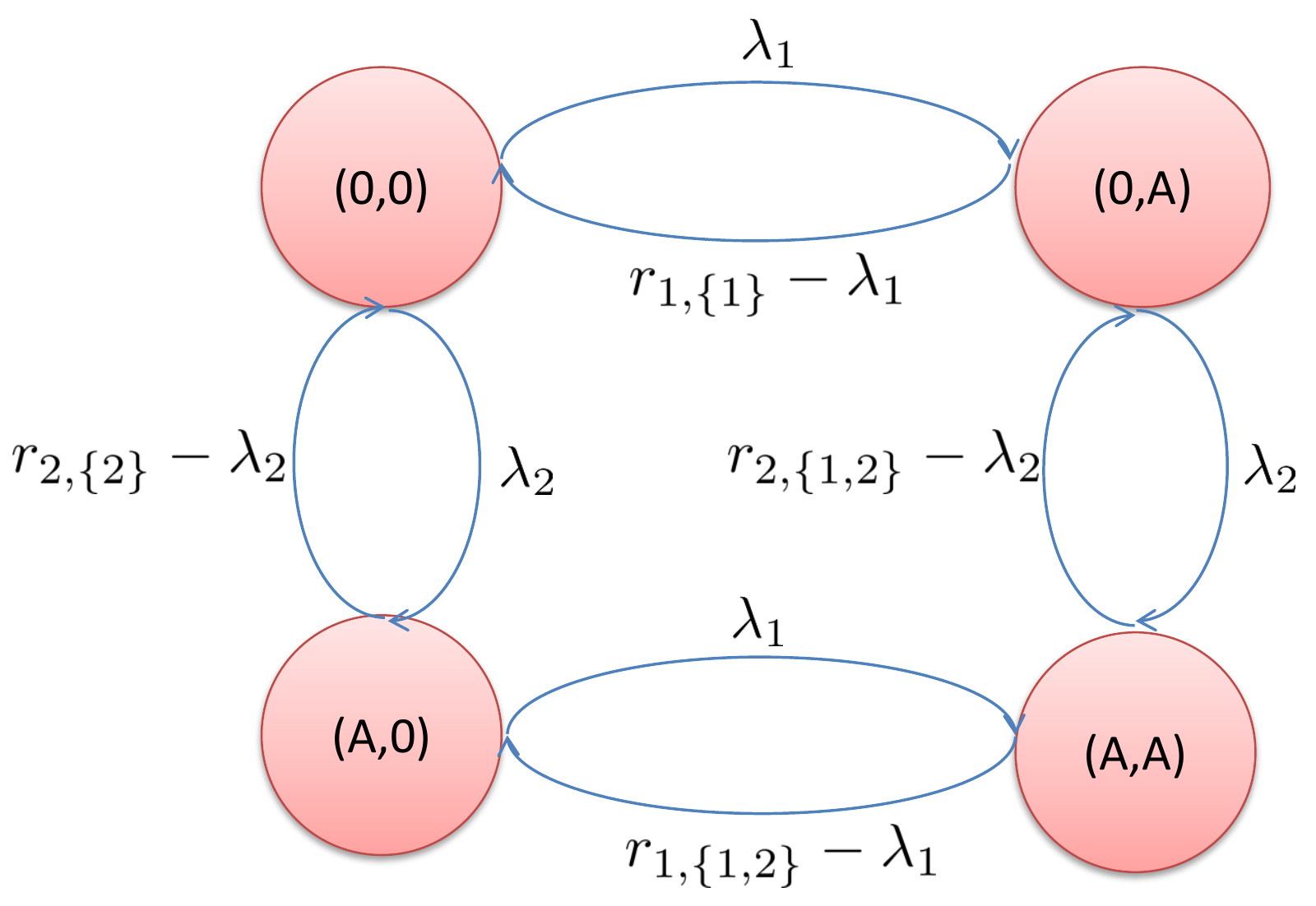}
\caption{The lumped CTMC for the 2-BTS interactive queueing model.}
\label{fig:LumpChain}
\end{figure}

The inter-group transitions can be modeled by an inter-group CTMC with $2^n$ states,
each corresponding to one group of active \BTSs\ $A$,
and hence is referred to as a {\em lumped state} in view of the original CTMC.
Two types of transitions can occur between the lumped states.
First, the inter-group CTMC may transit from a lumped state $B$ where queue $i$ is empty to
 another lumped state $A$ where queue $i$ becomes nonempty, i.e.,
\begin{align}   \label{eq:AB}
  \mathcalA=\mathcalB
  \cup \{i\}  \;\;\text{ and }\;\; i\notin\mathcalB.
\end{align}
The rate of such a transition is the rate that a packet arrives at queue $i$, $\lambda_i$.
Second, the inter-group CTMC may transit from a lumped state $A$ where queue $i$ has length 1 to another lumped state $B$ where queue $i$ becomes empty ($A$ and $B$ satisfy~\eqref{eq:AB}).
The rate of such a transition is the probability that the queue length $l=1$ times the service rate, and can be expressed as:
\begin{align}  \label{eq:5}
  p_i(1|A) r_{i,A}
  &= \left( 1 - \frac{\lambda_i}{r_{i,A}} \right) r_{i,A} \\
  &= r_{i,A} - \lambda_i.
\end{align}
The transition rates that completely describe the inter-group CTMC of the lumped states are expressed as:
\begin{align}\label{eq:ProbLump}
  \hat{Q}(\mathcalA,\mathcalB)
  =
  \begin{cases}
    \lambda_i, &\text{if}~\mathcalB=\mathcalA\cup\{i\}, i\notin A\\
    r_{i,\mathcalA}-\lambda_i, &\text{if}~\mathcalA=\mathcalB\cup \{i\}, i\notin B\\
    -\sum_{\mathcalC:\mathcalC\neq\mathcalA}\hat{Q}(\mathcalA,\mathcalC), &\text{if}~\mathcalA=\mathcalB\\
0,&~\text{otherwise}.
\end{cases}
\end{align}
As an example, the CTMC of the lumped states
is illustrated for the 2-BTS case 
in Fig.~\ref{fig:LumpChain}.
The steady-state distribution $p(\mathcalA)$ of the lumped states can be readily computed based on~\eqref{eq:ProbLump} using standard techniques in~\cite{Nel95Spinger}.
The key step therein is to invert a modified transition rate matrix.

\subsubsection{Average Delay} 



The average length of queue $i$ in the modified CTMC can be calculated as follows:
\begin{align} \label{eq:AvgDelay}
  \bar{l}_i
  &= \sum_{\bsl} l_i p(\bsl) \\
  &= \sum_{\bsl} l_i p(A_\bsl) \prod^n_{j=1} p_j(l_j|A_\bsl) \\
  &= \sum_A p(A) \sum_{\bsl: A_\bsl=A} l_i \prod^n_{j=1} p_j(l_j|A) \\
  &= \sum_A p(A) \sum_{l_i: i\in A} l_i p_i(l_i|A)
  \left( \prod_{j\in A \backslash\{i\} } \sum^\infty_{l_j=1} p_j(l_j|A)  \right)\\
  &= \sum_A p(A) \sum_{l_i: i\in A} l_i p_i(l_i|A) \\
  &= \sum_{A: i\in A} p(A) \sum^\infty_{l=1} l p_i(l|A) \\
  &= \sum_{\mathcalA:i\in\mathcalA}\frac{p(A) r_{i,\mathcalA}}{r_{i,\mathcalA}-\lambda_i}.
\end{align}
By Little's law, the average 
delay is given by $t_i=\bar{l}_i/\lambda_i$:
\begin{align}\label{eq:ApprDelay}
  t_i=\sum_{\mathcalA:i\in\mathcalA} \frac{p(A) r_{i,\mathcalA}}{(r_{i,\mathcalA}-\lambda_i)\lambda_i}.
\end{align}
\DG{This} is in general an approximation of 
\DG{the average delay} of the original CTMC.


\subsection{The Optimization Problem}
\label{s:opt2}

Using the approximate average delay
given by~\eqref{eq:ApprDelay}, we formulate the refined 
spectrum allocation problem as:
\begin{varsubequations}{P2}
\label{eq:SAP-Int}
\begin{align}
  \minimize\limits_{\xx,\rr,\ttv}\;\;
  &\sum_{i=1}^n \frac{\lambda_i}{\sum_{j=1}^{n}\lambda_j} t_i\label{eq:Obj-Int}\\
  \subjectto\;\;
  &t_i=\sum_{\mathcalA:i\in\mathcalA}
  \frac{p(\mathcalA)r_{i,\mathcalA}}{(r_{i,\mathcalA}-\lambda_i)\lambda_i},
  &\forall i\in N\\
  &r_{i,\mathcalA}=\sum_{\mathcalB\subset N}s_{i,\mathcalB\cap\mathcalA}x_{\mathcalB},
  \;\forall i\in N, \hspace{-2ex}&\forall\mathcalA\subset N\\
  &r_{i, N}>\lambda_i, &\forall i\in N\label{eq:Con2-Int}\\
  &x_{\mathcalB}\geq0, &\forall \mathcalB\subset N\label{eq:Con1-Int}\\
  &\sum_{\mathcalB\subset N}x_{\mathcalB}=1\label{eq:Con3-Int}
\end{align}%
\end{varsubequations}%
\DG{where $\ttv=[t_1,\dots,t_n]$ and $\rr=(r_{i,A})_{i\in N,A\subset N}$.}
Constraint \eqref{eq:Con2-Int} \DG{assures the stability of the CTMC within each lump state}.
\DG{It is not clear if Problem~\eqref{eq:SAP-Int} is convex due to the matrix inversion involved in calculating $p(A)$.  Neither can we establish a counterpart to Theorem~\ref{thm:KBTS} in this case, i.e., there is no guarantee that the optimal solution uses at most $n$ reuse patterns.}

\DG{Nonetheless, we use} a standard convex optimization algorithm to solve\DG{~\eqref{eq:SAP-Int}.} 
\DG{The} simulations in Section~\ref{sec:Sim} show that
\DG{the algorithm always converges} to the same solution regardless of the initial point.
\DG{The resulting solution} may divide the spectrum into \DG{a few} more than $n$ segments, \DG{but not by many}.

\subsection{\DG{Upper and Lower} Bounds \DG{on the Delay}}
\label{subsec:Pre}

\DG{Several upper and lower bounds can be obtained for both the {\em actual} average delay under the refined model and its approximation~\eqref{eq:ApprDelay}.  First, the conservative model's delay given by~\eqref{eq:IndDelay} is clearly an upper bound on the refined model's actual delay.
Using tools developed in~\cite{delcoigne2004modeling} for queues with time-varying link capacity,~\cite{bonald2004wireless} developed bounds for queueing systems with coupled processors.  The {\em first-degree bounds} therein are calculated assuming the best or worst service rates that decouple the queues.  In fact,~\eqref{eq:IndDelay} corresponds exactly to the first-degree upper bound.
}
The first-degree upper (lower) bound is usually loose in the light (heavy) traffic regime. 

Tighter {\em second-degree bounds} are also presented in~\cite{bonald2004wireless}. 
\DG{To lower (upper) bound the delay in cell} $i$, the
\DG{best (worst) rates are assumed} 
to determine the utilization 
\DG{in interfering cells}, i.e., the fraction of the time that
\DG{those \BTSs} 
are transmitting.
\DG{For example, to derive a} second-degree upper bound for
\DG{cell}
$i$,
\DG{every other BTS $j$ is} assumed to transmit at \DG{its worst} rate $r_{j, N}$. 
Denote $\mathcalA_i(t)$ and $\bar{\mathcalA}_i(t)$ as the set of active \BTSs\ exclusive of BTS $i$ at time $t$ in the original interactive queuing system and
under this assumption, respectively. It can be proved using a sample path argument that $\mathcalA_i(t)\subset\bar{\mathcalA}_i(t)$ 
\DG{at all time} $t$.
\DG{Thus}
the service rate at BTS $i$ under this assumption is always lower than the service rate in the original interactive queuing system. 
\DG{This gives the delay upper bound.}
\DG{To derive the} lower bound for
\DG{cell} 
$i$, all other \BTSs\ are assumed to transmit at the
\DG{best possible} 
rates, $r_{j,\{j\}},\,j\neq i$.

When assuming fixed rates at other \BTSs, the queue at BTS $i$ reduces to a queue with time-varying capacity. Processor sharing queues with time-varying capacity were studied in~\cite{delcoigne2004modeling}, which showed that the upper and lower bounds on residual work load in such queues can be obtained by considering the quasi-stationary regime and the fluid regime, respectively. In the quasi-stationary regime, the rest of the system evolves so slowly that BTS $i$ only sees the initial states of the other \BTSs. In the fluid regime the rest of the system evolves so quickly that BTS $i$ only sees the average interference.

Under the worse-case transmit rate assumption, the probability that BTS $j,\,j\neq i$ transmits is:
\begin{align}
\label{eq:UppActProb}
\bar{p}_j=\frac{\lambda_j}{r_{j, N}}.
\end{align}
The probability that the other $n-1$ \BTSs\ are in state $\bar{\mathcalA}_i$
is:
\begin{align}
\label{eq:UppStatProb}
\bar{\pi}_i(\bar{\mathcalA}_i)=\prod_{j\in\bar{\mathcalA}_i}\bar{p}_j \prod_{l\not\in\bar{\mathcalA}_i,l\neq i}(1-\bar{p}_l).
\end{align}
The second-degree upper bound is finally given by taking the expectation over the distribution of all possible states $\bar{\mathcalA}_i$:
\begin{align}
\label{eq:UppDelay}
\bar{t}_i=\sum_{\bar{\mathcalA}_i\subset( N\setminus \{i\})}
\bar{\pi}_i(\bar{\mathcalA}_i)\frac{1}{r_{i,\bar{\mathcalA}_i\cup\{i\}}-\lambda_i}.
\end{align}

Under the
\DG{best possible}
rate assumption, the probability of BTS $j,\,j\neq i$ being active is:
\begin{align}
\label{eq:LowActProb}
\underline{p}_j=\frac{\lambda_j}{r_{j,\{j\}}}.
\end{align}
The corresponding probability of state $\mathcal{\underline{A}}_i$ for the $n-1$ \BTSs\ except BTS $i$ is:
\begin{align}
\label{eq:LowStatProb}
\underline{\pi}_i(\underline{\mathcalA}_i)=\prod_{j\in\underline{\mathcalA}_i}\underline{p}_j \prod_{l\not\in\underline{\mathcalA}_i,l\neq i}(1-\underline{p}_l).
\end{align}
The second-degree lower bound is calculated using the average rate:
\begin{align}
\label{eq:LowDelay}
\underline{t}_i=\frac{1}{\sum_{\underline{\mathcalA}_i\subset( N\setminus\{i\})}
\underline{\pi}_i(\underline{\mathcalA}_i)r_{i,\underline{\mathcalA}_i\cup\{i\}}-\lambda_i}.
\end{align}

The accuracy of the refined approximation for the average delay developed in Section~\ref{subsec:RSAP} is guaranteed by the following theorem:

\begin{theorem}
\label{thm:Appr}
In a $n$-BTS interactive queueing system, the refined approximate mean packet sojourn time provided by~\eqref{eq:ApprDelay} is between the second-degree upper and lower bounds in~\eqref{eq:UppDelay} and~\eqref{eq:LowDelay}, i.e., $\underline{t}_i<t_i<\bar{t}_i$.
\end{theorem}

The proof of \textit{Theorem}~\ref{thm:Appr} is given in Appendix A.

\subsection{\DG{Throughput Optimality of the Proposed Schemes}}
\label{sec:Stable}


\DG{The throughput region of an interactive queueing system is given in~\cite{TasEph92TransAC} in the context of resource allocation through coordinated scheduling in the time domain.  Adapting the result to the current setting gives}
the throughput region of 
\DG{the} $n$-BTS network with 
given spectral efficiencies $(s_{i,B})_{i\in N,~B\subset N}$
\DG{as:} 
\begin{align}\label{eq:ThruptRegion}
  \Lambda=\big\{(\lambda_1,\dots,\lambda_n) \,\big|\,
\exists\,(x_B)_{B\subset N},\,\text{s.t.}\,r_i\geq\lambda_i,\forall i\in N \big\}
\end{align}
\DG{where $r_i$ is defined as in~\eqref{eq:r_i}.  That is,}
for any rate tuple in the interior of $\Lambda$, there exists \DG{a} spectrum allocation 
that stabilizes 
\DG{the} interactive \DG{queueing system}, 
\DG{whereas} for any rate tuples outside the region, there is no 
allocation that \DG{can} 
stabilize all the queues.

\begin{theorem} \label{thm:MaxThrupt}
  The proposed conservative and refined spectrum allocation schemes are both throughput optimal, namely, they both achieve the entire throughput region as given in~\eqref{eq:ThruptRegion}.
\end{theorem}
\begin{IEEEproof}
Problems~\eqref{eq:SAP-Ind} and~\eqref{eq:SAP-Int} both include the constraint, $r_i > \lambda_i~\forall i\in N$. Hence the feasible regions for the two proposed schemes are exactly the throughput region given by~\eqref{eq:ThruptRegion}.
\DG{Hence their throughput optimality follows.}
\end{IEEEproof}

The result in~\cite{TasEph92TransAC} only guarantees that for any rate tuple that can be stabilized, there exists a spectrum allocation $r_{i}\geq \lambda_i,~\forall i\in N$. However, there might be another spectrum allocations that also stabilizes the system, but have $r_{i}<\lambda_i$ at some of the \BTSs. The intuition is that although the worst case service rate is less than the traffic arrival rate at some queues, those queues will also operate at other states $A$ with less interference and higher rates. By Loynes theorem~\cite{Loy1962CPS}, the interactive queueing system will be stable, as long as the average service rate exceeds the packet arrival rate at each queue. Determining the stability of an interactive queueing system with fixed arrival rates $(\lambda_i)_{i\in N}$ and service rates $(r_{i,B})_{i\in N,~B\subset N}$ is more complicated. The result for two interactive queues is presented in~\cite{KnsMor2003JQS}.

\section{Numerical Results}
\label{sec:Sim}
Throughout this section, the noise PSD is $0.125\times 10^{-6}$ $\mu$W/Hz, the average packet size is $L=1$ Mbit and the total bandwidth is $W=20$ MHz in all simulations. In the homogenous setup, the path-loss exponent is 3, and the transmit PSD is 1 $\mu$W/Hz for all \BTSs. In the heterogenous setup,  the transmit PSD and the pathloss exponent of the macro BTS are 10 $\mu$W/Hz and 2.8, respectively; while the transmit PSD and the pathloss exponent for each pico BTS are 1 micro-watt/Hz and 3.4, respectively. The traffic distribution at each BTS is assumed to be proportional to its worst-case transmit rate, i.e., $\lambda_i\propto r_{i,N},~i\in N$. When the average load increases, $\lambda_1,\cdots,\lambda_N$ increase proportionally. However, the key conclusions apply to an arbitrary traffic distribution. The results for actual packet delays with service rates given by~\eqref{eq:ServiceRate}, are obtained by simulating the equivalent discrete time Markov chain (DTMC) of the CTMC~\eqref{eq:Prob} using the uniformization method~\cite{Ste09probability} for $10^5$ time intervals.

\subsection{One-Dimensional Example}
\label{subsec:1D}

\begin{figure}
\centering
\includegraphics[width=\mywidth]{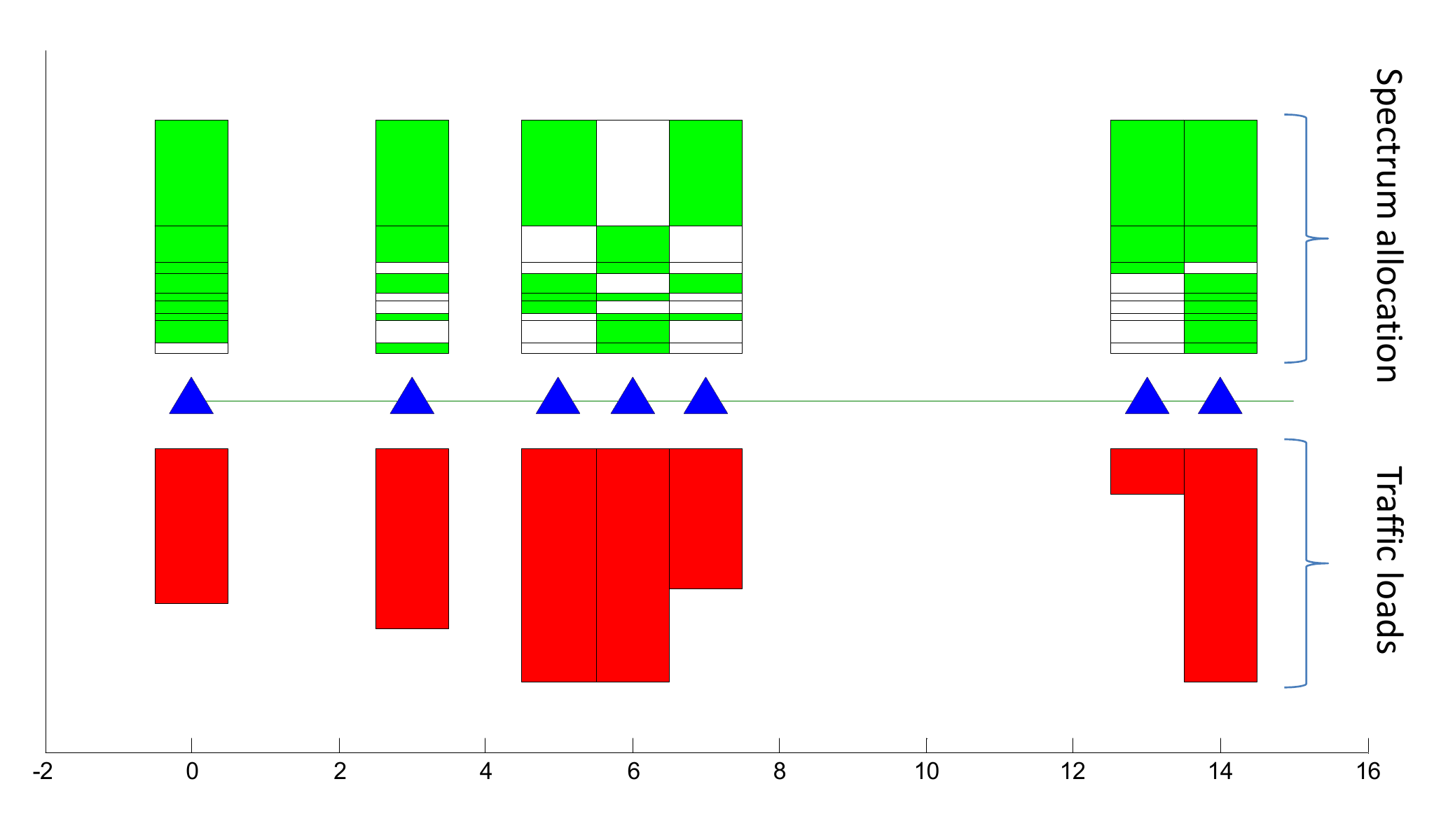}
\caption{One dimensional refined spectrum allocation.}
\label{fig:1D}
\end{figure}

We first present the one-dimensional example in Fig.~\ref{fig:1D} to \DG{demonstrate} 
that the proposed spectrum allocation scheme is topology- and traffic-aware.
Seven \BTSs, represented by the triangles, are randomly \DG{placed}
on a line segment.
The relative traffic load at each BTS is 
\DG{depicted} by the height of each rectangle below it,
\DG{whereas the spectrum allocated using the refined scheme is depicted using the rectangle above it.  A solid block depicts an active reuse pattern.}
\DG{Evidently,} both spatial reuse and local orthogonalization are
\DG{accomplished} 
by the 
allocation.
\DG{For example, the right most cell has much heavier traffic than its immediate neighbor, and is thus allocated much more spectrum.} 
Also, the spectrum allocated to the middle BTS is mostly orthogonal with those of the two close neighbors.

\subsection{Two-Dimensional Simulation Model}
\label{subsec:Sim}
\begin{figure}
\centering
\includegraphics[width=0.5\mywidth]{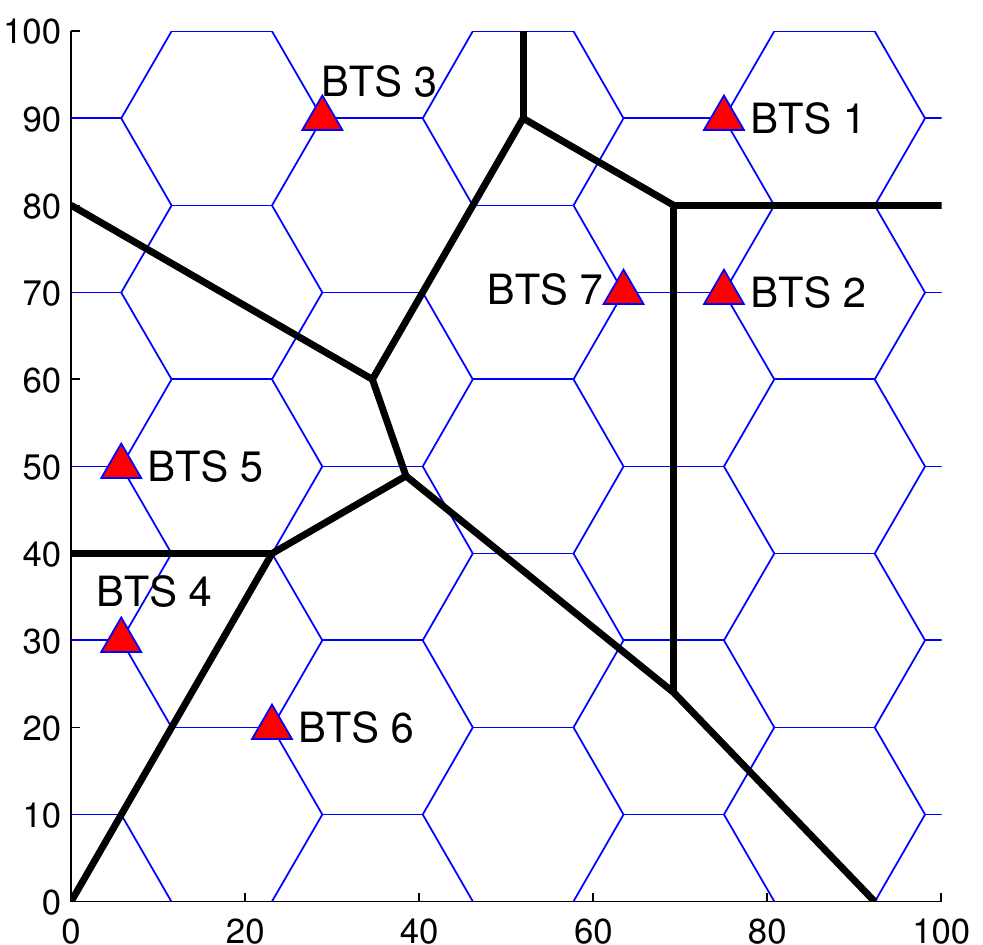}
\caption{The topology of the 7-pico BTS network.}
\label{fig:Topo}
\end{figure}

To illustrate the performance of the proposed conservative and refined
\DG{schemes}, 
we adopt the quantized HetNet model in~\cite{zhuang2012energy}. A $100\times100$~$\text{m}^2$ area is partitioned into hexagons, where the distance between nearest centers is $20$~m. In the simulation, 7 \BTSs\ are uniformly randomly dropped at the vertices of the hexagons. Each UE location is approximated by the center of its hexagon, and is assigned to its nearest BTS. The channel gain are determined using the standard path-loss model. \BZ{The topology of the network is shown in Fig.~\ref{fig:Topo}, where the locations of the BTSs are denoted by the triangles. The Voronoi region of each pico cell is also shown. The average spectral efficiency of BTS $i$ is calculated as the mean of the spectral efficiencies of the hexagons it serves.}

\subsection{Delay Comparison}
\label{subsec:DelayComp}
\begin{figure}
\centering
\includegraphics[width=\mywidth]{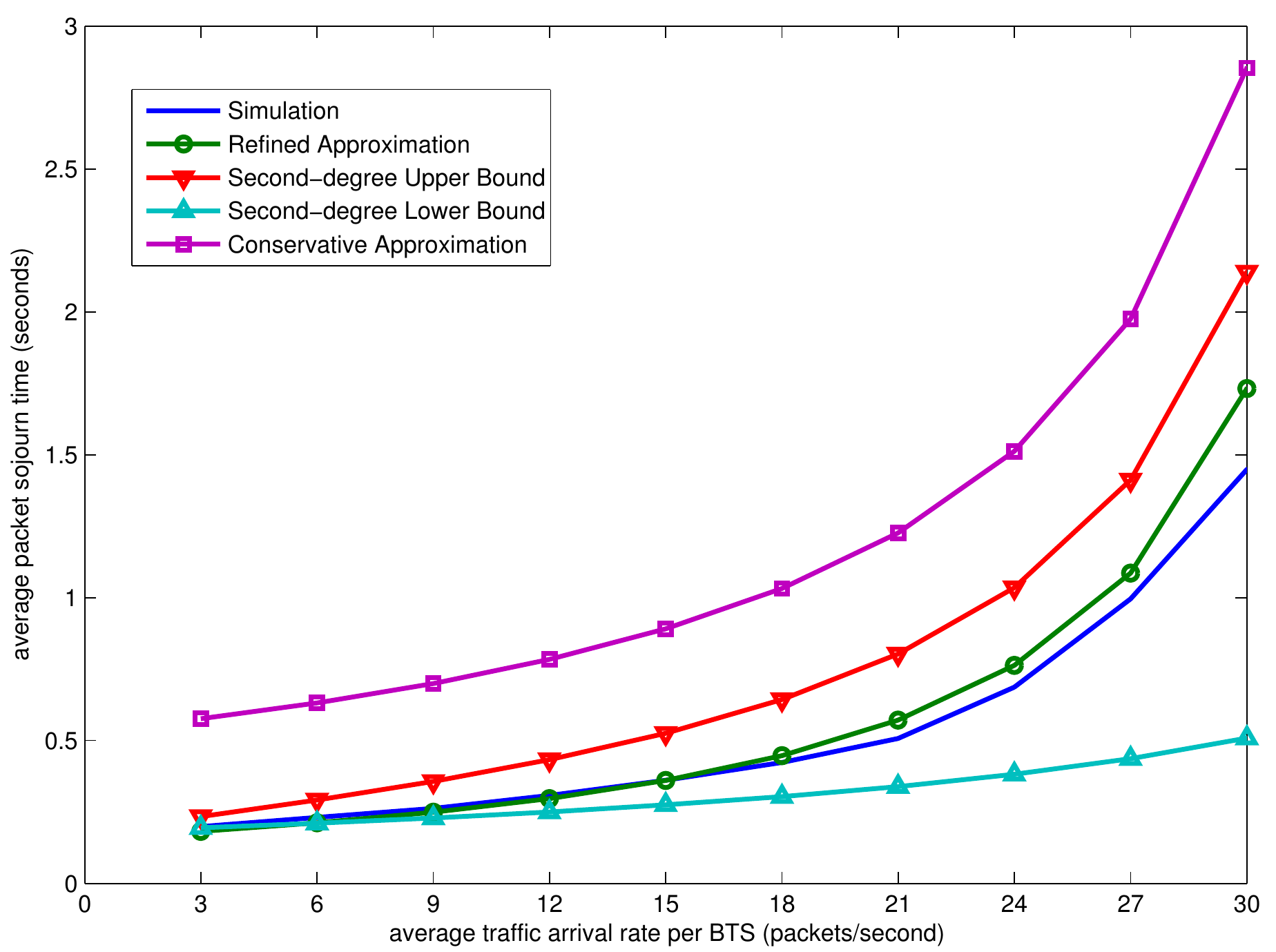}
\caption{Comparison of the conservative and refined approximations with the second-degree upper and lower bounds.}
\label{fig:Bounds}
\end{figure}

The delays using the refined and conservative approximations are compared with the second-degree upper and lower bounds in Fig.~\ref{fig:Bounds}.
The average delay versus average traffic arrival rate curves in Fig.~\ref{fig:Bounds} are based on the same spectrum allocation. The conservative approximation, also known as the first-degree upper bound, is coarser than the second-degree upper bound. As predicted by Theorem~\ref{thm:Appr}, the refined approximation is between the second-degree upper and lower bounds. In addition, the refined approximation is also quite accurate within the feasible region.

\begin{figure}
\centering
\includegraphics[width=\mywidth]{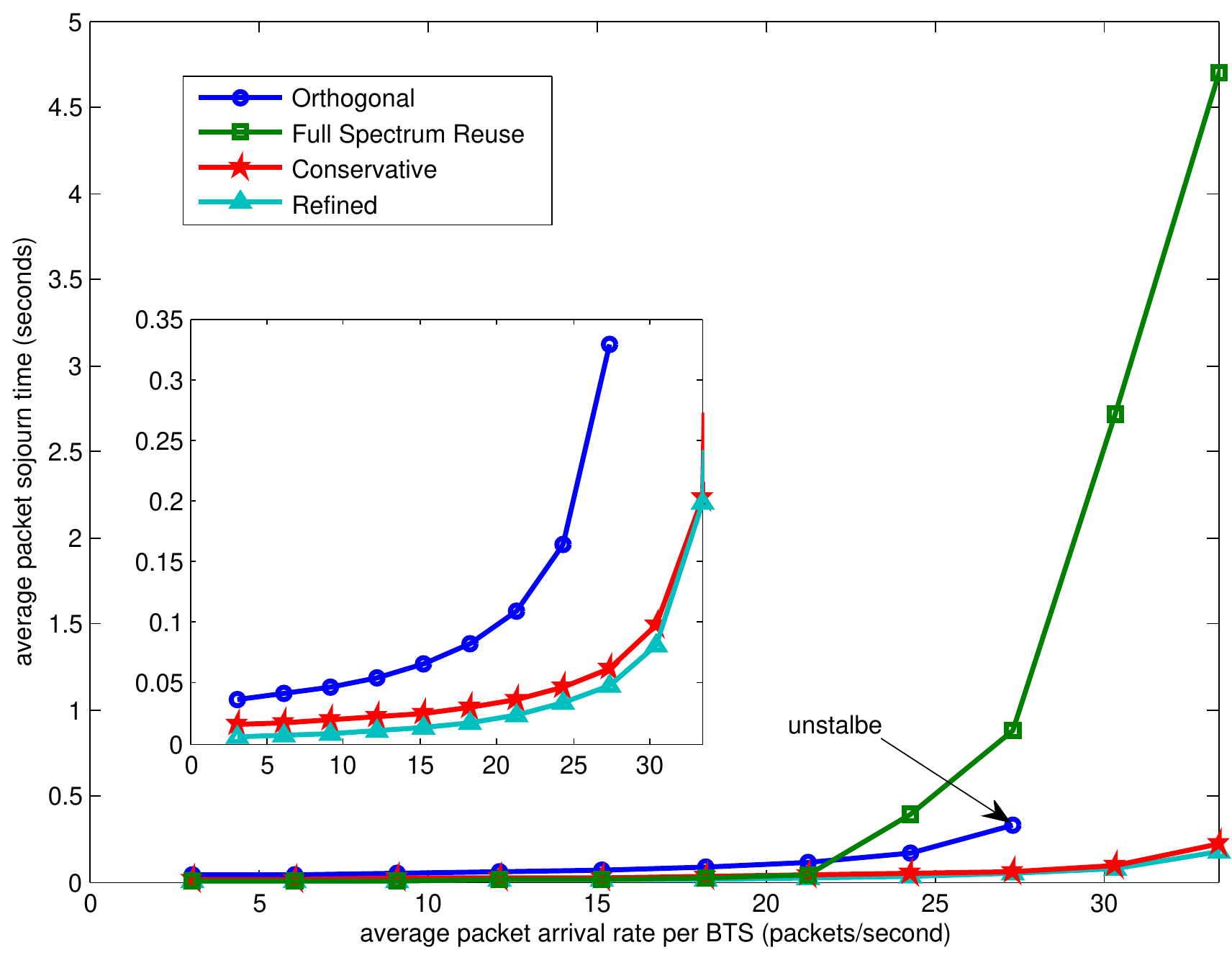}
\caption{Comparison of the conservative and refined allocations with the orthogonal and full-reuse allocations}
\label{fig:Delay}
\end{figure}

\begin{figure*}
\centering
\subfloat[The homogeneous setup with 7 pico BTSs.]{
\includegraphics[width=3in]{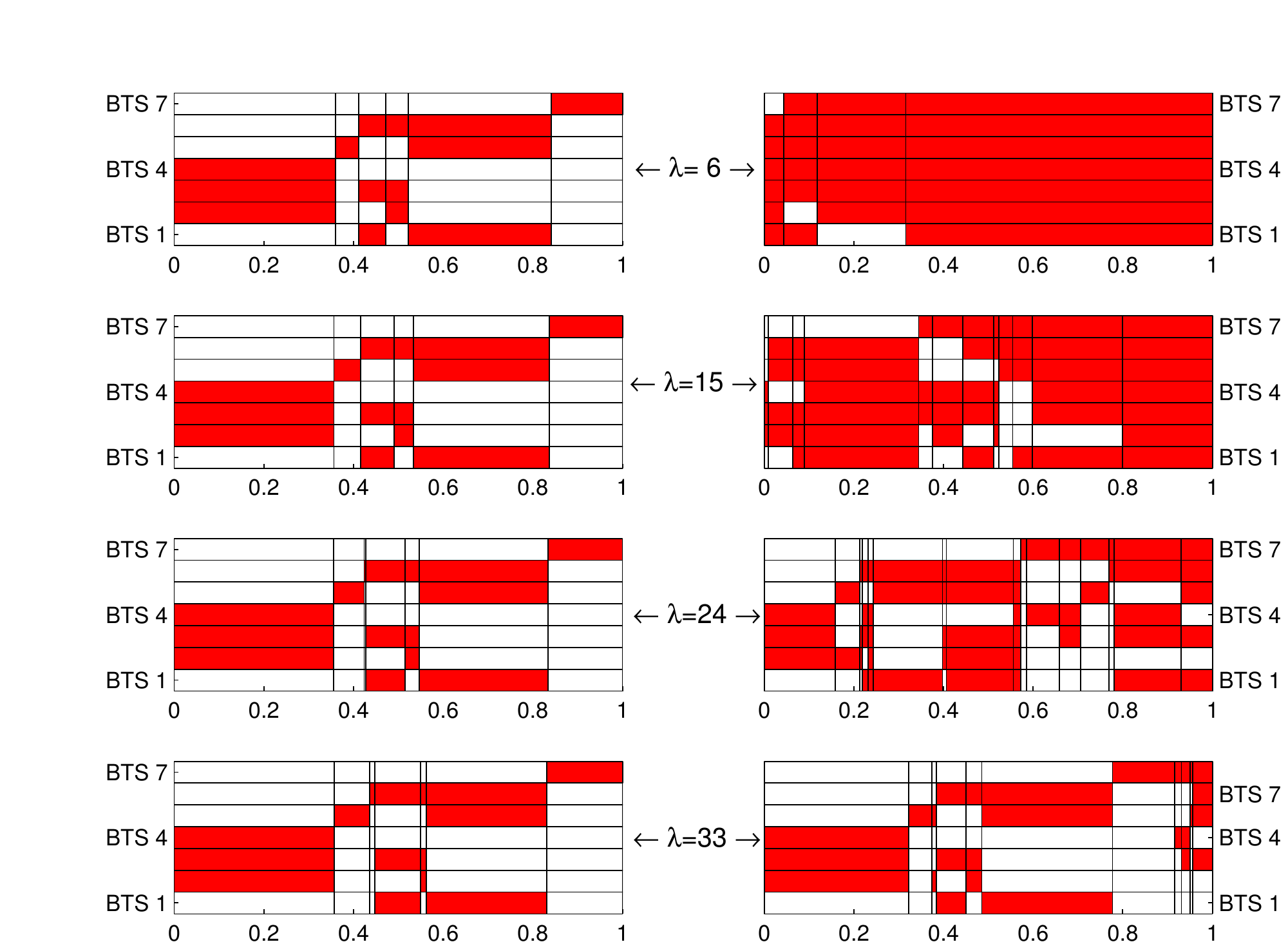}
\label{fig:Allocation}}
\hfil
\subfloat[The heterogeneous setup with 1 macro BTS and 7 pico BTSs]{
\includegraphics[width=3in]{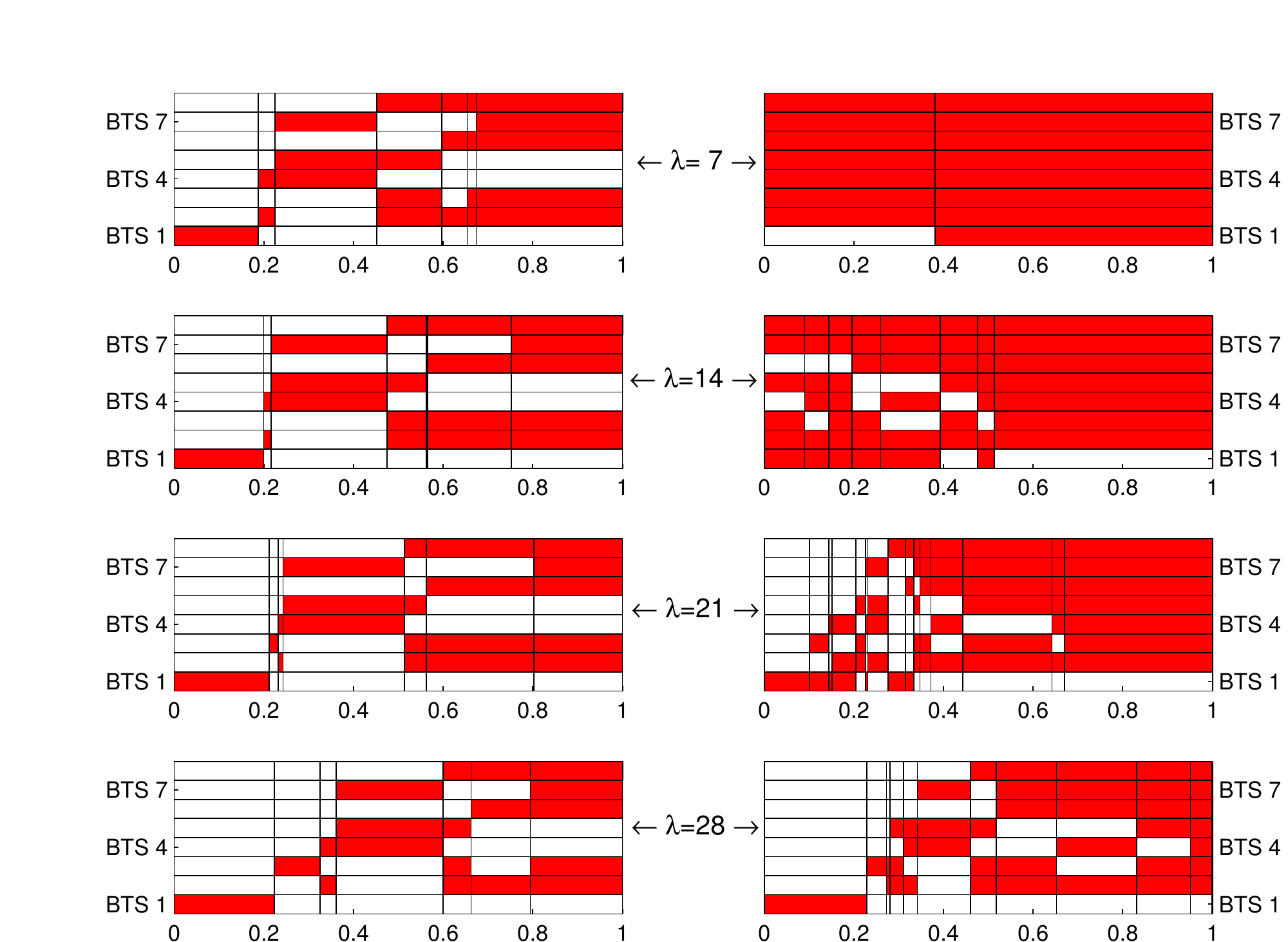}
\label{fig:AllocationHet}}
\caption{Optimal conservative (left) and refined (right) allocations with different packet arrival rates $\lambda$.}
\label{fig:Spectrum}
\end{figure*}

We compare the conservative and refined spectrum allocations with simpler orthogonal and full-reuse allocations for different traffic loads in Fig.~\ref{fig:Delay}.
The orthogonal allocation is the optimal one among all feasible orthogonal allocations.
The figure shows that the orthogonal allocation becomes unstable after the average packet arrival rate reaches 27 packets/second, suggesting this is the throughput region under the orthogonal allocation. The delay under the full-reuse allocation grows much faster after the average packet arrival rate rises above 21 packets/second. A general observation is that as the difference in spectral efficiency under the orthogonal and full-reuse allocations grows larger, \BZ{the delay under full-reuse also increases more rapidly with the traffic}. As suggested by the numerical results presented in Section~\ref{subsec:Uti}, the full-reuse allocation also becomes unstable at 27 pacekts/second.\footnote{The finite delay of the full-reuse allocation at arrival rate of 27 packets/second and even higher is due to finite simulation time.}
Hence, the proposed conservative and refined allocations achieve a larger throughput region than the other schemes. We can also observe from Fig.~\ref{fig:Delay} that the proposed conservative and refined allocations achieve significant gains in the heavy-traffic regime.


Since the delay under the full-reuse allocation becomes large in the heavy-traffic regime, the orthogonal, conservative and refined allocations are compared separately in the inlet in Fig.~\ref{fig:Delay}. Since the average sojourn time under the orthogonal allocation can be exactly calculated from~\eqref{eq:Obj-Ind}, the minimum of~\eqref{eq:SAP-Ind} given by the optimal conservative allocation is always no greater than the actual delay of the orthogonal allocation. Moreover, since the conservative approximation is an upper bound on the actual delay, the orthogonal allocation is always worse than the conservative allocation as shown in the inlet. The refined allocation always outperforms the conservative allocation due to the more accurate approximation of the actual delay. In the light-traffic regime, the refined allocation reduces the average delay by about 60\% compared to the conservative allocation; while both provide significant delay reduction compared to the orthogonal allocation. The advantage of using the refined allocation over the conservative allocation decreases as the traffic increases. In the heavy-traffic regime, the difference is negligible.


The optimal conservative and refined allocations are shown in Fig.~\ref{fig:Allocation} for different traffic loads.
The widths of the rectangles represent the \BZ{bandwidths of the active reuse patterns}. The solid ones in each row are the spectrum segments that are used by the corresponding BTS. The number of active reuse patterns employed by the conservative scheme is exactly the same as the number of cells (c.f., Theorem 1). In the light-traffic regime, the refined allocation is close to full-reuse, and as the traffic increases it becomes closer to the conservative allocation. This is because in the light-traffic regime all \BTSs\ are inactive most of the time, whereas in the heavy-traffic regime, all queues are mostly occupied, which perform similarly to $n$ independent M/M/1 queues with the worst-case service rates.

\subsection{Heterogenous Setup}
Fig.~\ref{fig:AllocationHet} illustrates the conservative and refined allocations for a HetNet with a single macro (large) cell and several pico (small) cells. The simulation setup is the same as described in Section~\ref{subsec:Sim}, with an additional macro BTS added at the center of the network. The proposed conservative and refined allocations are compared with the simple optimal orthogonal and full-reuse allocations at different traffic levels. The delay performance versus different loads looks similar to the one shown in Fig.~\ref{fig:Delay}, and is omitted.

As shown in Fig.~\ref{fig:AllocationHet}, the conservative allocation orthogonalizes the spectrum use between the macro BTS and the pico \BTSs\ in all traffic regimes. (BTS 1 is the macro BTS, whose spectrum allocation is shown at the bottom of each subplot.) This is because the macro BTS causes significant interference to the pico \BTSs\ under the backlogged interference assumption. In the refined allocation, the macro BTS can still share part of the spectrum with the pico \BTSs\ in the light-traffic regime. As the traffic load increases, an orthogonal spectrum allocation is observed between macro and pico tiers. Of course, the proposed spectrum allocation schemes can be applied to other more general HetNets, as long as the spectral efficiencies under different sharing combinations along with the traffic arrival rates can be obtained by a central controller.

%

\subsection{Utilization}
\label{subsec:Uti}

\begin{figure*}
\centering
\subfloat[average packet arrival rate = 3]{
\includegraphics[width =2in]{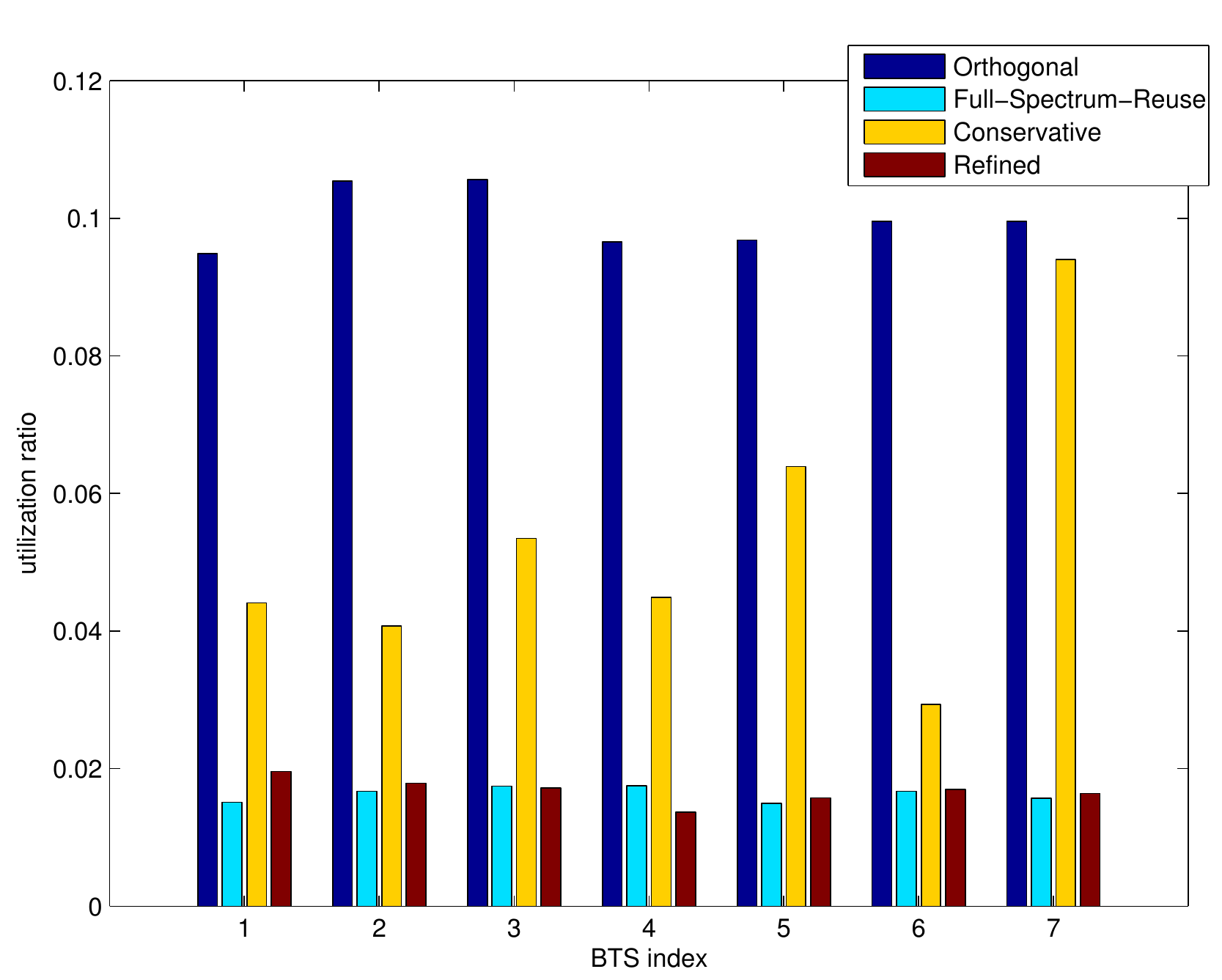}
\label{fig:bar1}}
\hfil
\subfloat[average packet arrival rate = 24]{
\includegraphics[width =2in]{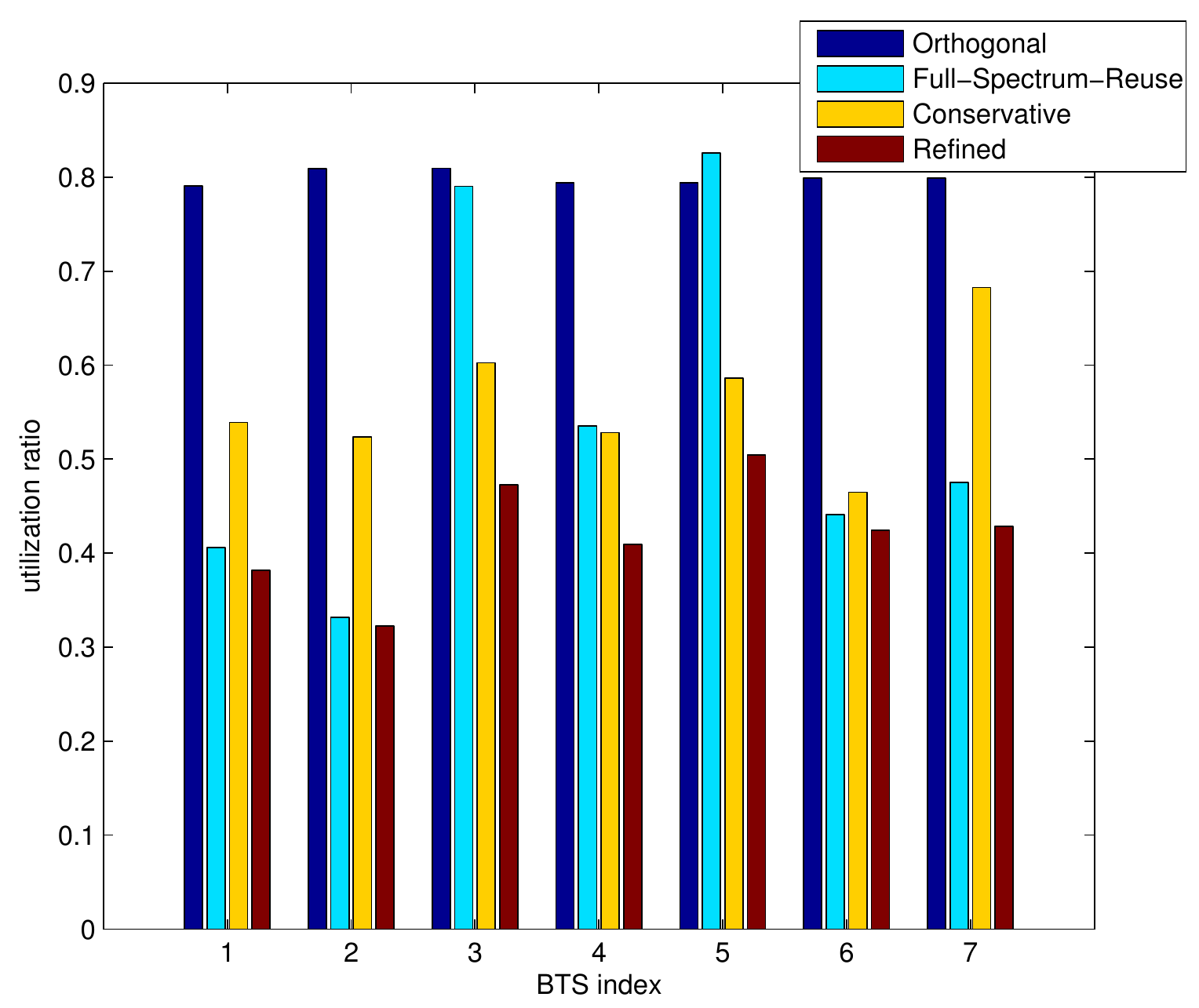}
\label{fig:bar8}}
\hfil
\subfloat[average packet arrival rate = 27]{
\includegraphics[width =2in]{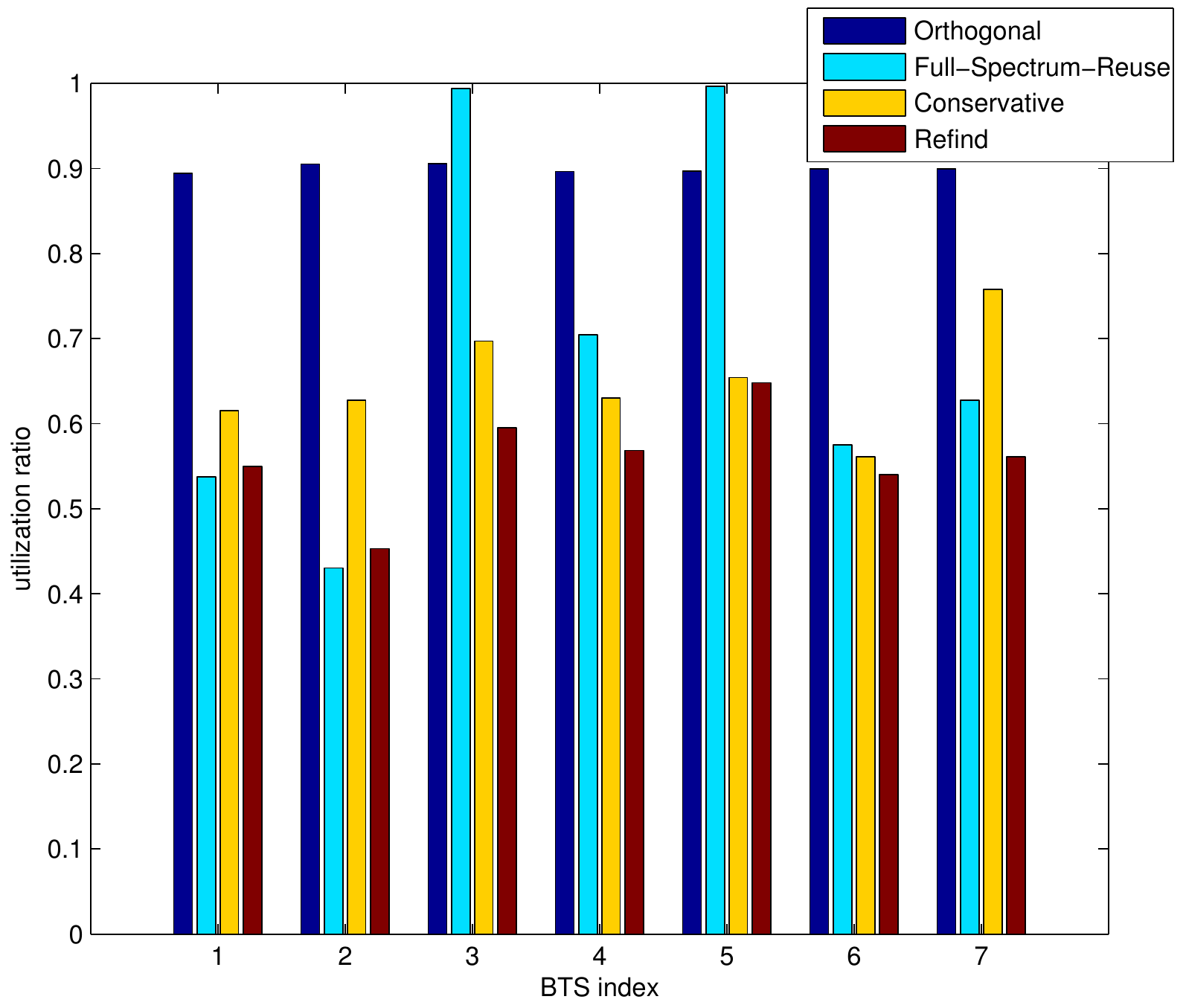}
\label{fig:bar9}}
\caption{BTS utilization ratios for different average packet arrival rates.}
\label{fig:UR}
\end{figure*}

We present the utilization ratio of each BTS under different traffic loads in Fig.~\ref{fig:UR}. In Fig.~\ref{fig:bar1}, where the traffic is very light, the refined allocation has very similar utilization as full-reuse. The utilization ratios are much higher under the conservative allocation due to the worst-case (smallest) transmit rate assumption. Since the \BTSs\ are active less than 10\% of the time, the orthogonal allocation is highly inefficient. Fig.~\ref{fig:bar8} presents the utilization ratios at an arrival rate where the full-reuse allocation starts to incur high delay as shown in Fig.~\ref{fig:Delay}. The conservative and refined allocations have much lower BTS utilization ratios compared to the full-reuse allocation, especially at BTS 3 and BTS 5. As the traffic demand increases, as shown in Fig.~\ref{fig:bar9}, \BTSs\ 3 and 5 become saturated under full-reuse, which also means the system becomes unstable. The conservative and refined allocations still maintain the stability of the queues and have much lower utilization ratios than the orthogonal allocation. As suggested in Fig.~\ref{fig:Delay}, if the traffic increases further to 30 packets/second, the orthogonal allocation also becomes unstable.
\subsection{Delay Distribution}
\label{subsec:DelayDis}

\begin{figure*}
\centering
\subfloat[average packet arrival rate = 3]{
\includegraphics[width =2in]{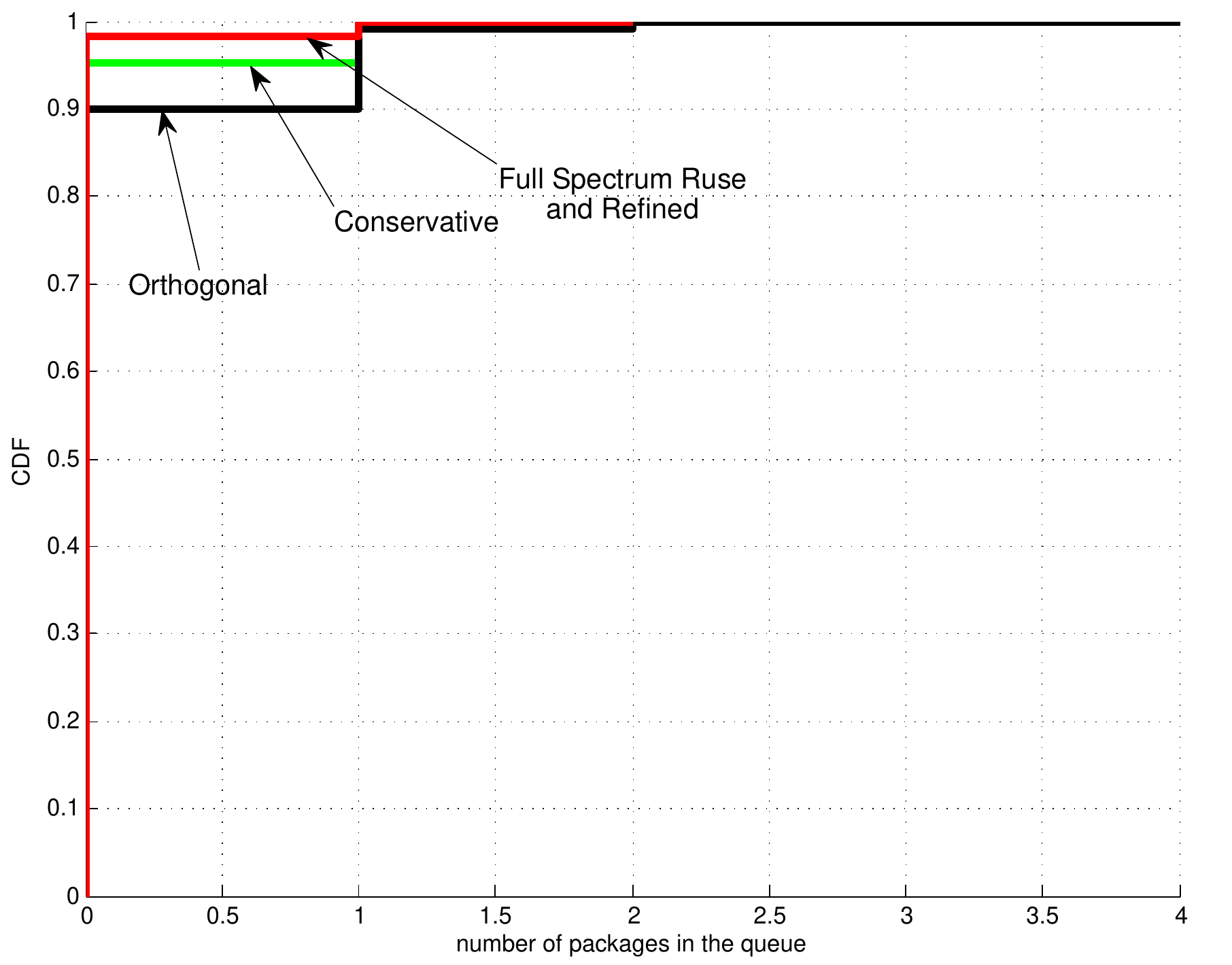}
\label{fig:CDF1}}
\hfil
\subfloat[average packet arrival rate = 24]{
\includegraphics[width =2in]{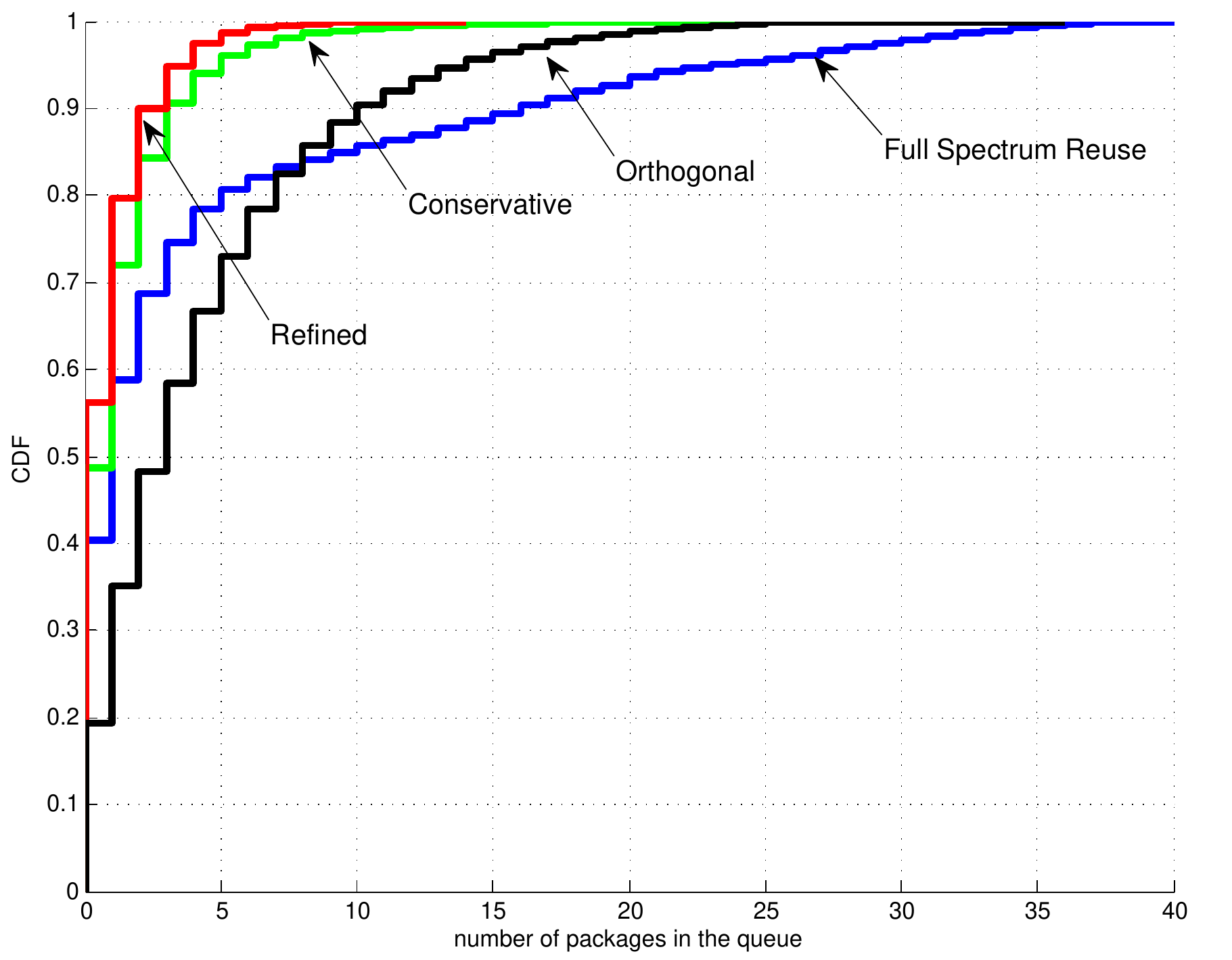}
\label{fig:CDF8}}
\hfil
\subfloat[average packet arrival rate =27]{
\includegraphics[width =2in]{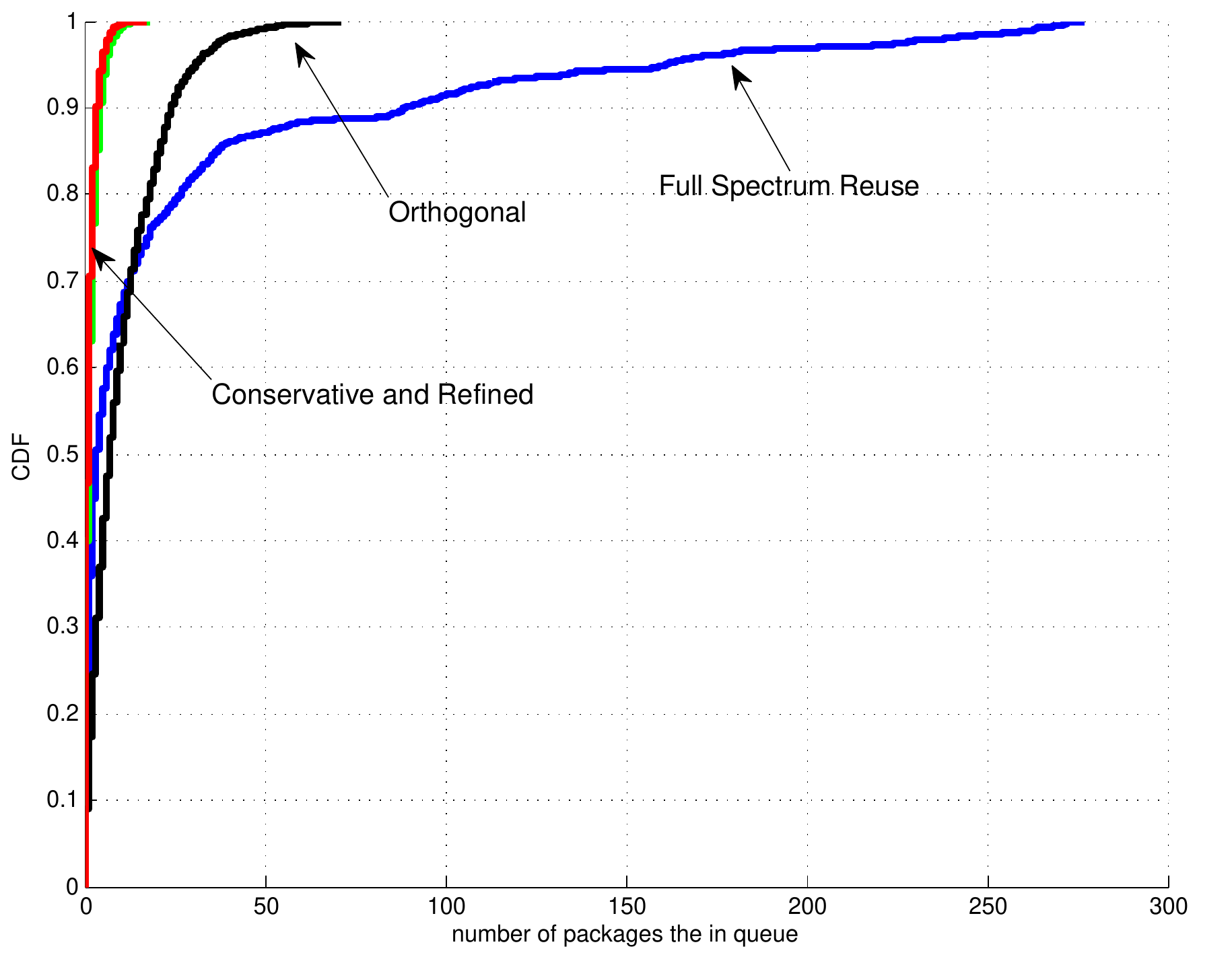}
\label{fig:CDF9}}
\caption{CDF's of number of packets in each queue for different average packet arrival rates.}
\label{fig:CDF}
\end{figure*}

The CDF of the number of packets in all queues for the entire network is shown in Fig.~\ref{fig:CDF}. In the light-traffic regime, as shown in Fig.~\ref{fig:CDF1}, the refined and full-reuse allocations have the smallest number of packets at all percentiles. The conservative allocation has more packets than those two, but fewer packets than the orthogonal allocation. The queues are empty 90\% of the time under all four allocations. At average traffic arrival rate of 24 packets/second, the refined and conservative allocations both have much fewer packets at all percentiles than the other two, as shown in Fig.~\ref{fig:CDF8}. The orthogonal allocation has a more balanced queue length compared with full-reuse due to fixed service rates.  At even higher traffic loads, as shown in Fig.~\ref{fig:CDF9}, the full-reuse allocation becomes unstable. The conservative and refined allocations have substantial advantages over the other two in this heavy-traffic regime.

\subsection{Power Control}
\label{subsec:Power}
The discussions so far have been based on the important assumption that the spectral efficiencies when shared by different combinations of \BTSs, i.e., $s_i(\mathcalA),~\forall i\in N,~\mathcalA\subset N$, are fixed. This assumption enables us to simplify the relation between spectrum allocation and flow level service rate. In fact, the service rate is a linear function of the allocated bandwidths under this fixed spectral efficiency assumption. This assumption is valid if all \BTSs\ transmit with fixed power spectral density.

However, in practice we may have a fixed total transmit power constraint at each BTS. Power control is not present in the current formulation. For example, the spectral efficiencies for the orthogonal allocation should be higher then shown, since each BTS would concentrate all the transmit power on its exclusive spectrum. The joint power and spectrum allocation problem in its general form can be formulated as optimizing a continuous power spectral density function at each BTS subject to a total transmit power constraint. The discretized version is proved to be NP-hard in~\cite{LuoZha08JSTSP}. The continuous version is also difficult.

We next take a simplified approach by alternatively updating the spectrum and power allocations. At the beginning, the spectral efficiencies are fixed assuming each BTS uniformly allocates its maximum transmit power across the entire spectrum. Then, we iterate the following steps:
\begin{enumerate}
\item Update the spectrum allocation $x_{\mathcalB},~\forall \mathcalB\subset N$ with the current $s_{i,\mathcalA},~\forall i\in N,~\mathcalA\subset N$ by solving the proposed spectrum allocation problems.

\item Update the spectral efficiencies $s_{i,\mathcalA},~\forall i\in N,~\mathcalA\subset N$ with the current $x_{\mathcalB},~\forall \mathcalB\subset N$, by letting each BTS uniformly allocate its maximum transmit power over the spectrum segments assigned to it.
\end{enumerate}

\begin{figure}
\centering
\includegraphics[width=\mywidth]{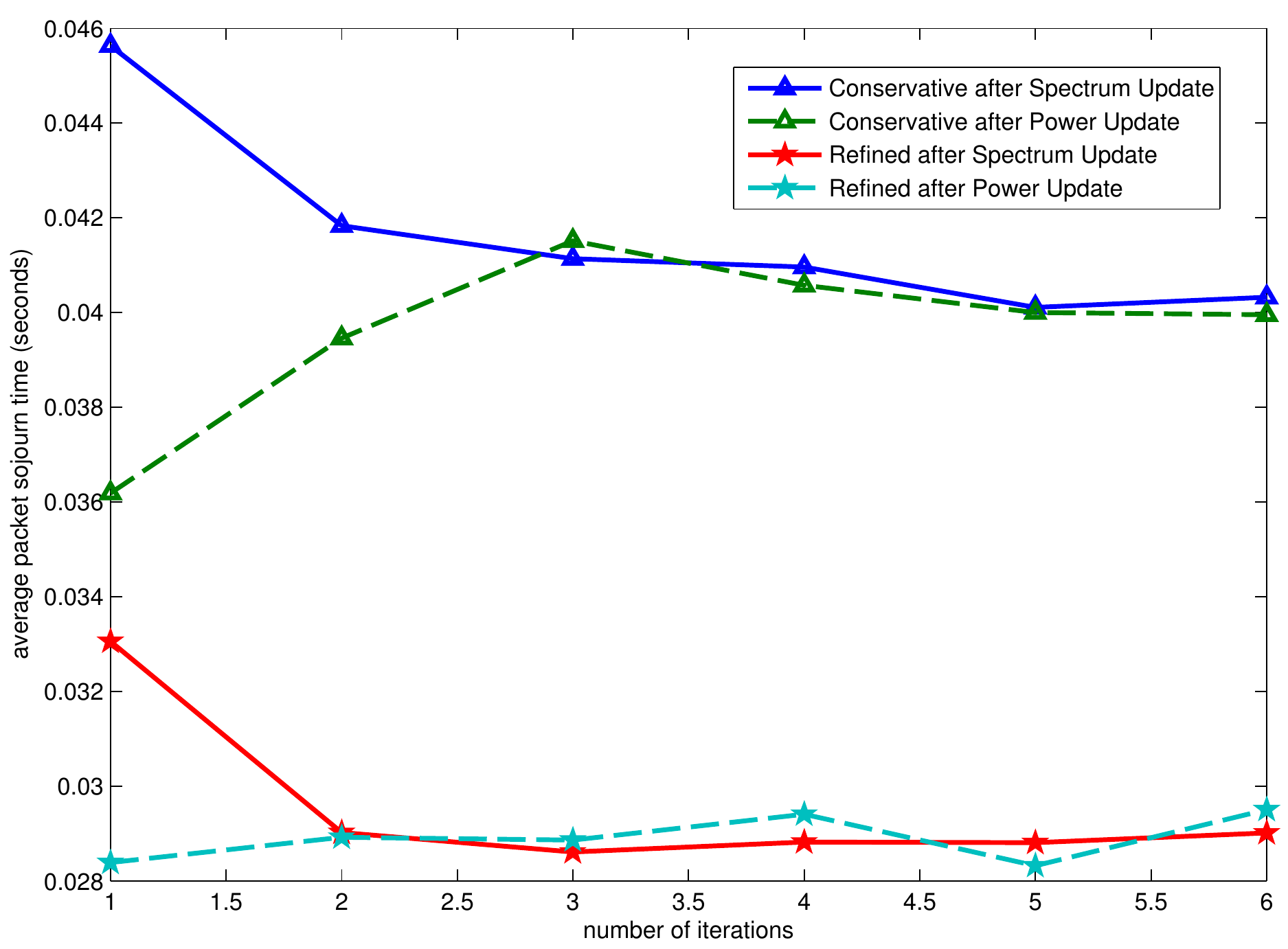}
\caption{Average packet sojourn time versus number of iterations.}
\label{fig:DelayIteration}
\end{figure}

The iterations continue until the spectrum allocation $\xx$ converges. (This is not guaranteed.) The average packet sojourn time after the spectrum allocation update and the spectral efficiency update at each iteration is shown in Fig.~\ref{fig:DelayIteration} for an average packet arrival rate per BTS of 24 packets/second.
 The figure shows that the delay performance converges very quickly for both conservative and refined allocations. The mean sojourn times for both allocations decrease substantially after the first spectral efficiency update. This is because at this average packet arrival rate, both allocations orthogonalize the spectrum use among neighboring \BTSs\ to some extent. (The small variations in these curves are due to the limited simulation time.)
This kind of convergence behavior can be expected in general, except for networks with very few \BTSs. This is because spatial reuse will occur in the conservative and refined allocations for relatively large networks even in the heavy-traffic regime. Since each BTS will use a fairly large amount of the spectrum, the spectral efficiencies will not change much after several iterations.

\section{Conclusion}
\label{sec:Con}

Traffic-\DG{driven} 
spectrum allocation performed over a slow timescale in a HetNet
has been studied.
\DG{Two efficient schemes based on optimization have been developed and shown to be highly effective.}
The proposed model and queueing analysis for densely deployed HetNets depart from the traditional single cell and regular hexagonal cellular network models.
Also, 
\DG{the solution fully adapts to} 
the topology of the network,
\DG{rather than} attempting to average over random realization \DG{using stochastic geometry}.

The conservative and refined allocations significantly reduce the average delay by exploiting the network topology and the different loads across \BTSs.
The problem formulation and results
\DG{can be generalized to arbitrary link performance measure and convex objective function, beyond}
the specific choice of Shannon spectral efficiency \DG{and queueing delay.  In particular, the formulation directly applies to the sum rate and weighted sum rate objectives.}

\DG{An important extension of the work is to develop a general framework for radio resource managment by incorporating user association, multiple antennas, and multiple radio access technologies (RATs).  The effort is well-matched to recent proposals for a cloud radio access network (C-RAN).  Ultimately, the goal is to also extended the framework to address large networks with hundreds to thousands of cells, thus pushing joint radio resource management to an unprecedented scale.}


\section*{Appendix A: Proof of Theorem~\ref{thm:Appr}}
\label{app:ProofThm2}
First we will prove $t_i<\bar{t}_i$. Both~\eqref{eq:ApprDelay} and~\eqref{eq:UppDelay} can be regarded as weighted sum of $\frac{r_{i,\mathcalA}}{(r_{i,\mathcalA}-\lambda_i)\lambda_i}$. The weight is the probability of the system being in state $\mathcalA$ under different approximation. Based on the worst-case rate assumption for deriving $\bar{t}_i$, we can form a similar lumped chain as~\eqref{eq:ProbLump} with the transition rate matrix:
\begin{align}\label{eq:ProbUpp}
  \bar{Q}(\mathcalA,\mathcalB)
  =
  \begin{cases}
    \lambda_i,&\text{if}~\mathcalB=\mathcalA\cup\{i\}, i\notin A\\
    (r_{i, N}-\lambda_i),&\text{if}~\mathcalA=\mathcalB\cup\{i\}, i\notin B\\
    -\sum_{\mathcalC\neq\mathcalA}\bar{Q}(\mathcalA,\mathcalC),&\text{if}~\mathcalA=\mathcalB\\
    0,&\text{otherwise}.
  \end{cases}
\end{align}
The lumped chains~\eqref{eq:ProbLump} and~\eqref{eq:ProbUpp} can be interpreted as the CTMC of the following Markov process: There are $n$ interactive queues each with capacity 1. Packet arrivals at queue $i$ follow a Poisson process with rate $\lambda_i$. The size of each packet is independently exponentially distributed with unit mean. The service rates are given by the corresponding rate matrices of~\eqref{eq:ProbLump} and~\eqref{eq:ProbUpp}, which are state dependent. If a packet arrives at an empty queue, it is immediately served. Otherwise, the packet in service will be discarded and the newly arrived packet immediately starts being served. Denote $S$ and $\bar{S}$ respectively as the corresponding systems with probability transition matrix given by~\eqref{eq:ProbLump} and~\eqref{eq:ProbUpp}. Let $\mathcalA(t)$ and $\bar{\mathcalA}(t)$ be the sets of active queues at time $t$. Denote the residual loads (in bits) by $\bsl(t)$ and $\bar{\bsl}(t)$, where $\bsl(t),~\bar{\bsl}(t)\in\mathds{R}_+^n$ and the $i$th elements $l_i(t)$ and $\bar{l}_i(t)$ are the residual loads in queue $i$ of the two systems. Using a sample path argument, we can show:
\begin{align}
\label{eq:SamplePath}
\begin{aligned}
&\mathcalA(t)\subset \bar{\mathcalA}(t)\\
&\bsl(t)\leq \bar{\bsl}(t)
\end{aligned}
\end{align}
at any time instance $t$. The inequality $\bsl(t)\leq \bar{\bsl}(t)$ is element-wise, i.e., $\l_i(t)\leq \l_i(t),~i=1, \dots, n$.

Assuming both systems evolve under the same packet arrival realization, we can prove \eqref{eq:SamplePath} by induction. At $t=0$, $\mathcalA(0)=\bar{\mathcalA}(0)=\emptyset$ and $\bsl(0)=\bar{\bsl}(0)=0$. Assume~\eqref{eq:SamplePath} is true at $t=\tau>0$. Then,~\eqref{eq:SamplePath} still holds before any arrival or departure happens at $t=\tau+\delta$. This is because $\mathcalA(t)\subset \bar{\mathcalA}(t),~\forall t\in[\tau,\tau+\delta)$ implies the service rate of any queue $i$ in $S$ is larger than the service rate of queue $i$ in $\bar{S}$ within this time interval. At $t=\tau+\delta$, we have two cases.

\begin{itemize}
\setlength{\itemindent}{+.5in}
    \item[Case 1.] An arrival happens at queue $j$. Since both systems follows the same sample path of the arrivals, we still have $\mathcalA(\tau+\delta)\subset \mathcalA(\tau+\delta)$. The residual load at other queues does not change from $\tau+\delta_-$ to $\tau+\delta$. At queue $j$, $\bsl_j(\tau+\delta)=\bar{\bsl}_j(\tau+\delta)$.
    \item[Case 2.] A departure happens at queue $j$. Since $\bsl(t)\leq \bar{\bsl}(t),~\forall t\in[\tau,\tau+\delta)$ and the service rate at each queue in $S$ is no less than the service rate at the same queue in $\bar{S}$, this departure must be in $S$ (It could be the case that a departure also happens at queue $j$ in $\bar{S}$ at $t=\tau+\delta$.). It immediately implies~\eqref{eq:SamplePath} at $t=\tau+\delta$.
\end{itemize}
Because the Markov processes in both systems are ergodic, we have:
\begin{align}
\label{eq:TimeAvg}
\begin{aligned}
&t_i=\lim_{T\rightarrow\infty}\frac{1}{T}\int_0^T f_i\left(\mathcalA(t)\right)d\mathcalA(t)\\
&\bar{t}_i=\lim_{T\rightarrow\infty}\frac{1}{T}\int_O^T f_i\left(\bar{\mathcalA}(t)\right)d\bar{\mathcalA}(t),
\end{aligned}
\end{align}
where the function $f_i(\mathcalA)$ is:
\begin{align}
\nonumber
f_i(\mathcalA)=\left\{
\begin{aligned}
&\frac{r_{i,\mathcalA}}{(r_{i,\mathcalA}-\lambda_i)\lambda_i}~&\text{if}~i\in\mathcalA\\
&0~&\text{otherwise}
\end{aligned}
\right.
\end{align}
It is easy to see that if $\mathcalA\subset\mathcalB$ then $f_i(\mathcalA)\leq f_i(\mathcalB)$. Applying this and~\eqref{eq:SamplePath} to~\eqref{eq:TimeAvg}, we can obtain $t_i< \bar{t}_i$, where the strict inequality is due to ergodicity.

To prove $t_i> \underline{t}_i$, we first introduce an intermediate variable $t'_i$:
\begin{align}
\label{eq:LowApprDelay}
t'_i=\frac{1}{\sum_{\mathcalA: i\in\mathcalA} P(\mathcalA)\frac{r_{i,\mathcalA}}{\lambda_i}r_{i,\mathcalA}-\lambda_i}.
\end{align}
According to~\eqref{eq:ProbLump}, it is easy to check $\sum_{\mathcalA: i\in\mathcalA} P(\mathcalA)\frac{r_{i,\mathcalA}}{\lambda_i}=1$. By Jensen's inequality, we immediately have $t_i\geq t'_i$. Hence, we only need to show $\hat{\underline{t}}_i>\underline{t}_i$. Again we can form a lumped chain with the highest service rate assumption, whose transition rate matrix is:
\begin{align}\label{eq:ProbLow}
  \underline{Q}(\mathcalA,\mathcalB)
  =
  \begin{cases}
    \lambda_i, &\text{if}~\mathcalB=\mathcalA\cup\{i\}, i\notin A\\
    \left(r_{i,\{i\}}-\lambda_i\right), &\text{if}~\mathcalA=\mathcalB\cup\{i\}, i\notin B\\
    -\sum_{\mathcalC:\mathcalC\neq\mathcalA}\underline{Q}(\mathcalA,\mathcalC),  &\text{if}~\mathcalA=\mathcalB\\
0, &\text{otherwise}.
\end{cases}
\end{align}
Using a similar sample path argument we can prove the expected service rate in~\eqref{eq:LowApprDelay} is less than the expected service rate in~\eqref{eq:LowDelay}, which directly implies $t'_i>\underline{t}_i$.



\begin{thebibliography}{10}

\bibitem{ZhuGuoHon14GLOBE}
B.~Zhuang, D.~Guo, and M.~L. Honig, ``Traffic driven resource allocation in
  heterogenous wireless networks,'' in {\em Proc.\ IEEE GLOBECOM},
  pp.~1546--1551, Dec 2014.

\bibitem{lei2007novel}
H.~Lei, L.~Zhang, X.~Zhang, and D.~Yang, ``A novel multi-cell {OFDMA} system
  structure using fractional frequency reuse,'' in {\em Proc.\ IEEE PIMRC}, pp.~1--5, Sept 2007.

\bibitem{andrews2014what}
J.~G. Andrews, S.~Buzzi, W.~Choi, S.~V. Hanly, A.~Lozano, A.~C.~K. Soong, and
  J.~C. Zhang, ``What will {5G} be?,'' {\em {IEEE} J. Sel. Areas Commun.},
  vol.~32, no.~6, pp.~1065--1082, 2014.

\bibitem{liu2014dense}
J.~Liu, W.~Xiao, and A.~C.~K. Soong, {\em Design and Deployment of Small Cell
  Networks}, ch.~Dense Networks of Small Cells.
\newblock Cambridge University Press, 2014.

\bibitem{stolyar2008self-organizing}
A.~L. Stolyar and H.~Viswanathan, ``Self-organizing dynamic fractional
  frequency reuse in {OFDMA} systems,'' in {\em Proc.\ IEEE INFOCOM}, Apr.
  2008.

\bibitem{chang2009multicell}
R.~Chang, Z.~Tao, J.~Zhang, and C.-C. Kuo, ``Multicell {OFDMA} downlink
  resource allocation using a graphic framework,'' {\em {IEEE} Trans. Veh.
  Technol.}, vol.~58, pp.~3494--3507, Sept 2009.

\bibitem{ali2009dynamic}
S.~H. Ali and V.~C.~M. Leung, ``Dynamic frequency allocation in fractional
  frequency reused {OFDMA} networks,'' {\em {IEEE} Trans. Wireless Commun.},
  vol.~8, pp.~4286--4295, Aug. 2009.

\bibitem{madan2010cell}
R.~Madan, J.~Borran, A.~Sampath, N.~Bhushan, A.~Khandekar, and T.~Ji, ``Cell
  association and interference coordination in heterogeneous {LTE-A} cellular
  networks,'' {\em {IEEE} J. Sel. Areas Commun.}, vol.~28, no.~9,
  pp.~1479--1489, 2010.

\bibitem{liao2014base}
W.-C. Liao, M.~Hong, Y.-F. Liu, and Z.-Q. Luo, ``Base station activation and
  linear transceiver design for optimal resource management in heterogeneous
  networks,'' {\em {IEEE} Trans. Signal Process.}, vol.~62, no.~15,
  pp.~3939--3952, 2014.

\bibitem{huang2009joint}
J.~Huang, V.~Subramanian, R.~Agrawal, and R.~Berry, ``Joint scheduling and
  resource allocation in uplink {OFDM} systems for broadband wireless access
  networks,'' {\em {IEEE} J. Sel. Areas Commun.}, vol.~27, pp.~226--234,
  February 2009.

\bibitem{fooladivanda2013joint}
D.~Fooladivanda and C.~Rosenberg, ``Joint resource allocation and user
  association for heterogeneous wireless cellular networks,'' {\em {IEEE}
  Trans. Wireless Commun.}, vol.~12, no.~1, pp.~248--257, 2013.

\bibitem{lim2014energy-efficient}
G.~Lim, C.~Xiong, L.~J. Cimini, and G.~Y. Li, ``Energy-efficient resource
  allocation for {OFDMA}-based multi-{RAT} networks,'' {\em {IEEE} Trans.
  Wireless Commun.}, vol.~13, no.~5, pp.~2696--2705, 2014.

\bibitem{shen2014distributed}
K.~Shen and W.~Yu, ``Distributed pricing-based user association for downlink
  heterogeneous cellular networks,'' {\em {IEEE} J. Sel. Areas Commun.},
  vol.~32, no.~6, pp.~1100--1113, 2014.

\bibitem{deb2014algorithms}
S.~Deb, P.~Monogioudis, J.~Miernik, and J.~P. Seymour, ``Algorithms for
  enhanced inter-cell interference coordination {(eICIC)} in {LTE} {HetNets},''
  {\em {IEEE/ACM} Trans. Netw.}, vol.~22, no.~1, pp.~137--150, 2014.

\bibitem{dhillon2013load-aware}
H.~Dhillon, R.~K. Ganti, and J.~G. Andrews, ``Load-aware modeling and analysis
  of heterogeneous cellular networks,'' {\em {IEEE} Trans. Wireless Commun.},
  vol.~12, no.~4, pp.~1666--1677, 2013.

\bibitem{shojaeifard2014unified}
A.~Shojaeifard, K.~A. Hamdi, E.~Alsusa, D.~K.~C. So, and J.~Tang, ``A unified
  model for the design and analysis of spatially-correlated load-aware
  {HetNets},'' {\em {IEEE} Trans. Commun.}, vol.~62, no.~11, pp.~4110--4125,
  2014.

\bibitem{bonald2004wireless}
T.~Bonald, S.~Borst, N.~Hegde, and A.~Prouti{\'e}re, ``Wireless data
  performance in multi-cell scenarios,'' in {\em Proceedings of the Joint
  International Conference on Measurement and Modeling of Computer Systems},
  SIGMETRICS/Performance, (New York, NY, USA), pp.~378--380, ACM, 2004.

\bibitem{rengarajan2008architecture}
B.~Rengarajan and G.~de~Veciana, ``Architecture and abstractions for
  environment and traffic aware system-level coordination of wireless networks:
  the downlink case,'' in {\em Proc.\ IEEE INFOCOM}, 2008.

\bibitem{KuaUtsDot2014arxiv}
Q.~Kuang, W.~Utschick, and A.~Dotzler, ``Optimal joint user association and
  resource allocation in heterogeneous networks via sparsity pursuit,'' {\em
  http://arxiv.org/abs/1408.5091}, 2014.

\bibitem{Hav2013Springer}
M.~Haviv, {\em Queues: A Course in Queueing Theory}, vol.~191.
\newblock Springer, 2013.

\bibitem{Nel95Spinger}
R.~Nelson, {\em Probability, stochastic processes, and queueing theory: the
  mathematics of computer performance modeling}.
\newblock Springer, 1995.

\bibitem{BerDim1999nonlinear}
D.~P. Bertsekas, {\em Nonlinear programming}.
\newblock Athena Scientific, 1999.

\bibitem{Egg69Convexity}
H.~G. Eggleston, {\em Convexity}.
\newblock No.~47, Cambridge University Press Archive, 1958.

\bibitem{BerDim97LP}
D.~Bertsimas and J.~N. Tsitsiklis, {\em Introduction to linear optimization},
  vol.~6.
\newblock Athena Scientific Belmont, MA, 1997.

\bibitem{FayIas1979Springer}
G.~Fayolle and R.~Iasnogorodski, ``Two coupled processors: the reduction to a
  riemann-hilbert problem,'' {\em Zeitschrift f{\"u}r
  Wahrscheinlichkeitstheorie und verwandte Gebiete}, vol.~47, no.~3,
  pp.~325--351, 1979.

\bibitem{AdaWes1993AAP}
I.~J. B.~F. Adan, J.~Wessels, and W.~H.~M. Zijm, ``A compensation approach for
  two-dimensional markov processes,'' {\em Advances in Applied Probability},
  vol.~25, pp.~783--817, Dec. 1993.

\bibitem{CohBox2000Elsevier}
J.~W. Cohen and O.~J. Boxma, {\em Boundary value problems in queueing system
  analysis}.
\newblock Elsevier, 2000.

\bibitem{delcoigne2004modeling}
F.~Delcoigne, A.~Prouti¨¨re, and G.~R¨¦gni¨¦, ``Modeling integration of
  streaming and data traffic,'' {\em Performance Evaluation}, vol.~55, no.~3-4,
  pp.~185--209, 2004.

\bibitem{ZhuGuoHon14arxiv}
B.~Zhuang, D.~Guo, and M.~L. Honig, ``Traffic-driven spectrum allocation in
  heterogeneous networks,'' {\em http://arxiv.org/abs/1408.6011}, 2014.

\bibitem{TasEph92TransAC}
L.~Tassiulas and A.~Ephremides, ``Stability properties of constrained queueing
  systems and scheduling policies for maximum throughput in multihop radio
  networks,'' {\em {IEEE} Trans. Autom. Control}, vol.~37, pp.~1936--1948, Dec.
  1992.

\bibitem{Loy1962CPS}
R.~M. Loynes, ``The stability of a queue with non-independent inter-arrival and
  service times,'' in {\em Cambridge Philos. Soc}, vol.~58, pp.~497--520,
  Cambridge Univ Press, 1962.

\bibitem{KnsMor2003JQS}
C.~Knessl and J.~A. Morrison, ``Heavy traffic analysis of two coupled
  processors,'' {\em Queueing Systems}, vol.~43, no.~3, pp.~173--220, 2003.

\bibitem{Ste09probability}
W.~J. Stewart, {\em Probability, Markov chains, queues, and simulation: the
  mathematical basis of performance modeling}.
\newblock Princeton University Press, 2009.

\bibitem{zhuang2012energy}
B.~Zhuang, M.~L. Honig, and D.~Guo, ``Energy management of dense wireless
  heterogeneous networks over slow timescales,'' in {\em Proc.\ Allerton Conf.\
  Commun., Control, \& Computing}, 2012.

\bibitem{LuoZha08JSTSP}
Z.-Q. Luo and S.~Zhang, ``Dynamic spectrum management: Complexity and
  duality,'' {\em {IEEE} J. Sel. Topics Signal Process.}, vol.~2, pp.~57--73,
  Feb 2008.

\end{thebibliography}

\begin{IEEEbiography}[{\includegraphics[width=1in,height=1.25in,clip,keepaspectratio]{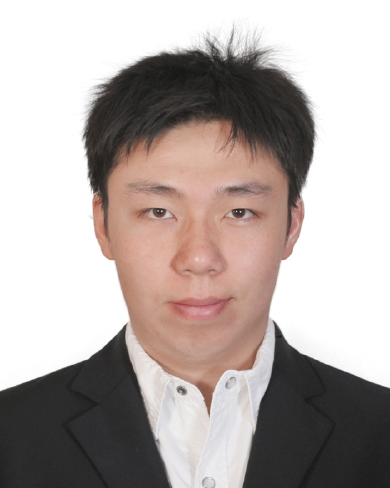}}]{Binnan Zhuang}
received his B.S. degree from Electronic Engineering Department of Tsinghua University in July 2009 the M.S. degree EECS department of Northwestern University. He is pursuing his Ph.D. degree at EECS department of Northwestern University. His research interests include wireless communications, communication network and network optimization. His current research focus on interference and resource management in heterogeneous networks.
\end{IEEEbiography}
\begin{IEEEbiography}[{\includegraphics[width=1in,height=1.25in,clip,keepaspectratio]{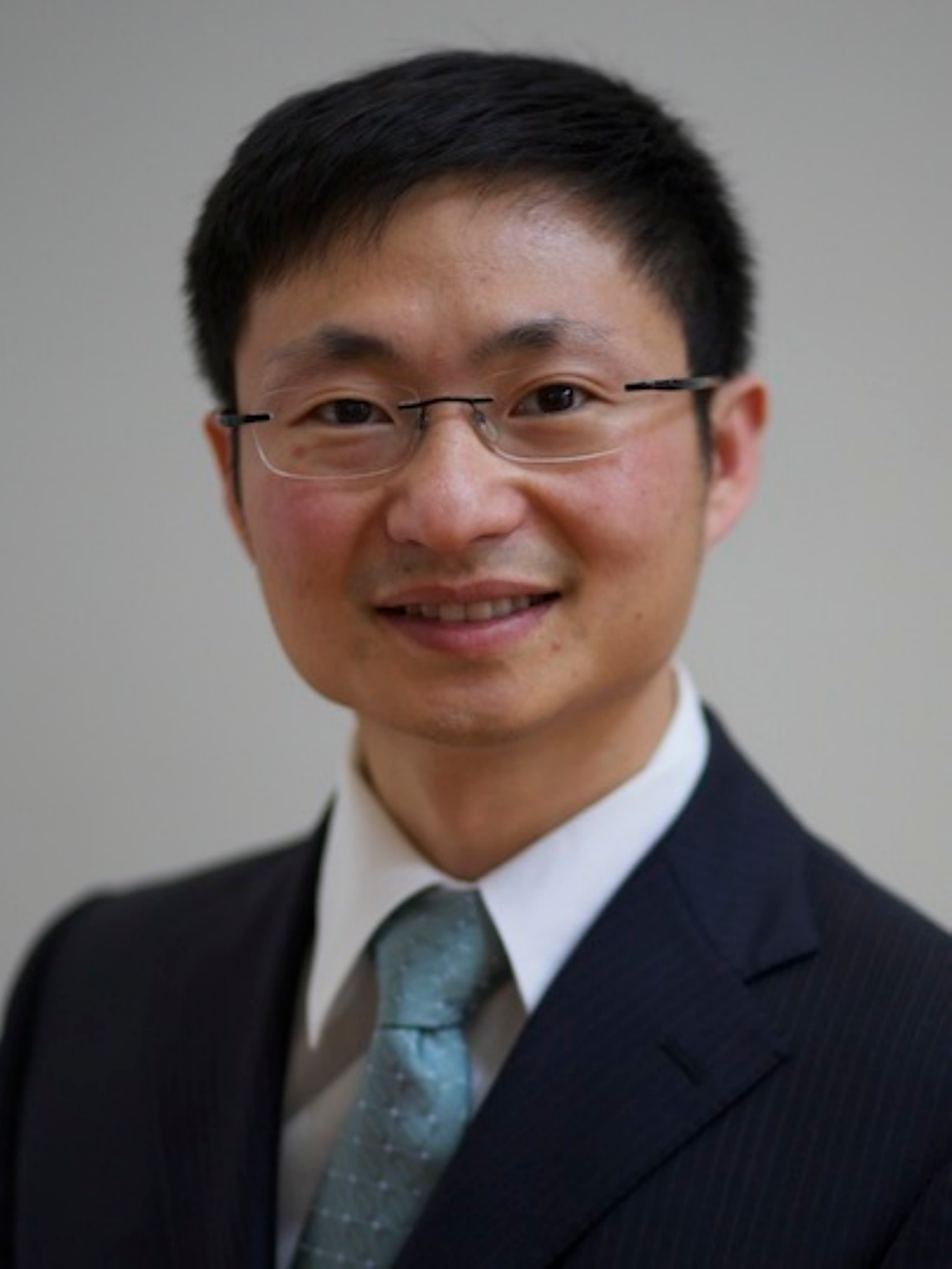}}]{Dongning Guo}
(S'97-M'05-SM'11) received the B.Eng.\ degree from the University of Science and Technology of China, Hefei, China, the M.Eng.\ degree from the National University of Singapore, Singapore, and the M.A.\ and Ph.D.\ degrees from Princeton University, Princeton, NJ, USA. In 2004, he joined the faculty of Northwestern University, Evanston, IL, USA, where he is currently an Associate Professor in the Department of Electrical Engineering and Computer Science. From 1998 to 1999, he was an R\&D Engineer in the Center for Wireless Communications, Singapore. He has been an Associate Editor of IEEE Transactions on Information Theory and a Guest Editor of a Special Issue of IEEE Journal on Selected Areas in Communications. He is an Editor of Foundations and Trends in Communications and Information Theory. Dr.~Guo received the Huber and Suhner Best Student Paper Award in the International Zurich Seminar on Broadband Communications in 2000 and is the corecipient of the IEEE Marconi Prize Paper Award in Wireless Communications in 2010. He is also the recipient of the National Science Foundation Faculty Early Career Development (CAREER) Award in 2007. His research interests include information theory, communications, and networking.
\end{IEEEbiography}
\begin{IEEEbiography}[{\includegraphics[width=1in,height=1.25in,clip,keepaspectratio]{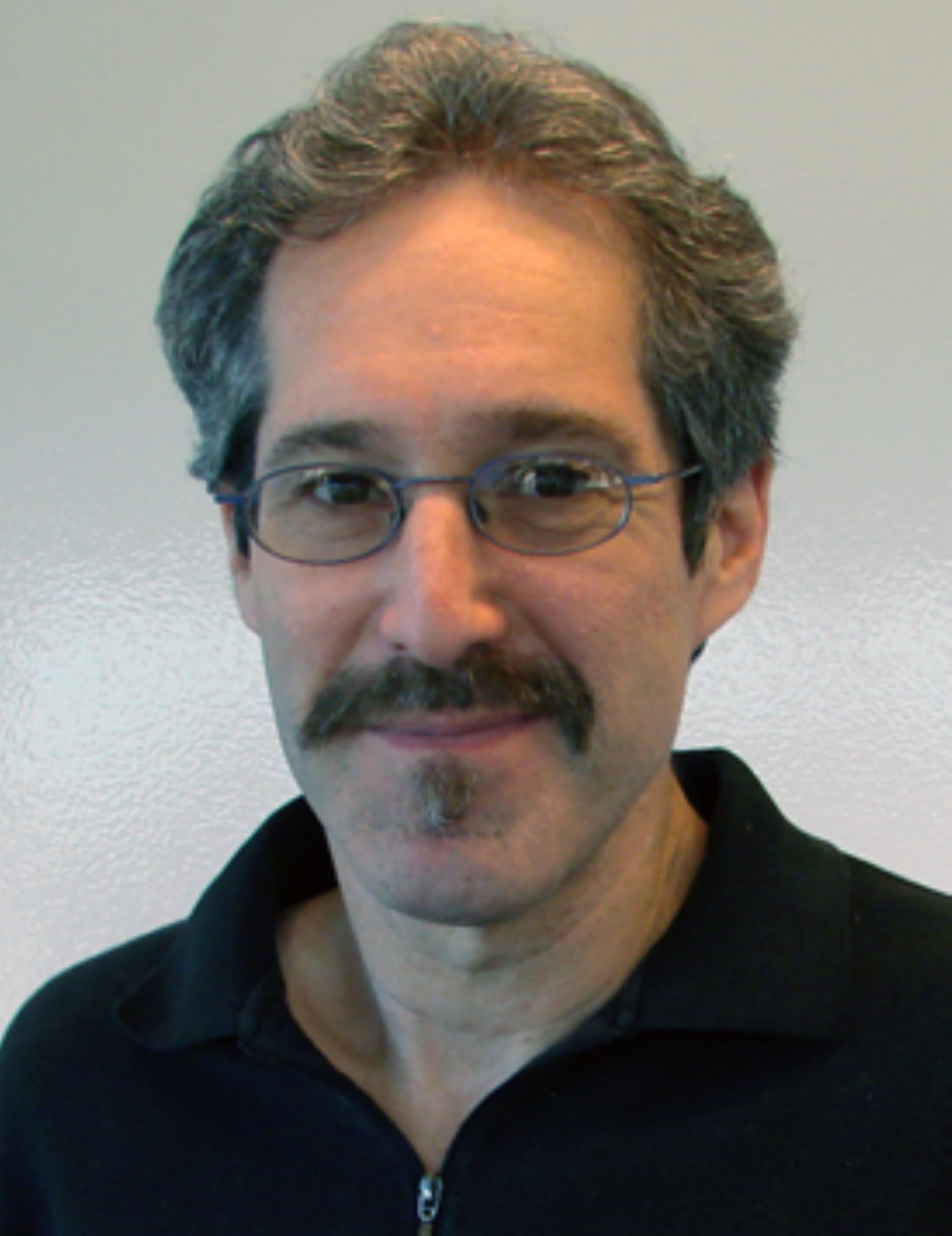}}]{Michael L. Honig}
(S'80-M'81-SM'92-F'97) received
the B.S. degree in electrical engineering from
Stanford University in 1977, and the M.S. and Ph.D.
degrees in electrical engineering from the University
of California, Berkeley, in 1978 and 1981, respectively.
He subsequently joined Bell Laboratories
in Holmdel, NJ, where he worked on local area
networks and voiceband data transmission. In 1983
he joined the Systems Principles Research Division
at Bellcore, where he worked on Digital Subscriber
Lines and wireless communications. Since the Fall
of 1994, he has been with Northwestern University where he is a Professor
in the Department of Electrical and Computer Engineering. He has held
visiting scholar positions at the Naval Research Laboratory (San Diego),
the University of California, Berkeley, the University of Sydney, Princeton
University, and the Technical University of Munich. He has also worked as a
freelance trombonist.

Dr. Honig has served as an Editor for the IEEE Transactions on Information
Theory and the IEEE Transactions on Communications,
and as Guest Editor for several journals. He has also served
as a member of the Board of Governors for the Information
Theory Society. He is the recipient of a Humboldt Research
Award for Senior U.S. Scientists, and the corecipient of the 2002 IEEE
Communications Society and Information Theory Society Joint Paper Award
and the 2010 IEEE Marconi Prize Paper Award.
\end{IEEEbiography}

\end{document}